\documentclass[longauth]{aa} 
\usepackage{colortbl}
\usepackage{rotating}
\usepackage{caption}
\usepackage{tabularx}
\usepackage{graphicx}
\usepackage{txfonts}
\usepackage{comment}
\usepackage{float}
\usepackage{multirow}
\usepackage{subcaption}
\usepackage[colorlinks=true, linkcolor=blue, citecolor=blue]{hyperref}
\usepackage[nameinlink,noabbrev]{cleveref}
\usepackage{fancyhdr}
\newcommand{\mh}{~Mpc$\,h^{-1}$ }
\newcommand{\mhp}{~Mpc$\,h^{-1}$}

\begin{document} 

   \title{The J-PAS survey: The effect of photometric redshift errors on cosmic voids}

   \author{J.A. Mansour$^1$, L. J. Liivamägi$^1$, A. Tamm$^1$, J. Laur$^1$, R. Abramo$^{2,3}$, E. Tempel$^{1,4}$, R. Kipper$^1$, A. Hernán-Caballero$^5$, V. Marra$^{14, 15, 16}$, J.Alcaniz$^6$, N. Benitez$^7$, S. Bonoli$^{8,9}$, S. Carneiro$^{10}$, J. Cenarro$^5$, D. Cristóbal-Hornillos$^5$, R. Dupke$^{6, 10, 11}$, A. Ederoclite$^5$, C. Hernández-Monteagudo$^{17, 18}$ , C. López-Sanjuan$^5$, A. Marín-Franch$^5$, C. M. de Oliveira$^{12}$, M. Moles$^{5, 7}$, L. Sodré Jr$^{12}$, K. Taylor$^{13}$, J. Varela$^5$, H. Vázquez Ramió$^5$
          }
   \authorrunning{Short Author List}
   \institute{$^1$  Tartu Observatory, University of Tartu, Observatooriumi 1, 61602 Tõravere, Estonia \\
   $^2$ Instituto de Física, Universidade de Sãoo Paulo, Rua do Matãoo 1371, 05508-090 Sãoo Paulo, Brazil \\
   $^3$ Departamento de Astronomia, Instituto de Astronomia, Geofísica e Ciências Atmosféricas, Universidade de São Paulo, São Paulo, Brazil \\
   $^4$ Estonian Academy of Sciences, Kohtu 6, 10130 Tallinn, Estonia \\
   $^5$ Centro de Estudios de Física del Cosmos de Aragón (CEFCA), Plaza San Juan, 1, E-44001 Teruel, Spain \\
   $^6$ Observatório Nacional, Ministério da Ciencia, Tecnologia, Inovação e Comunicações, Rua General José Cristino, 77, São Cristóvão, 20921-400, Rio de Janeiro, Brazil \\
   $^7$ Instituto de Astrofísica de Andalucía (CSIC), P.O. Box 3004, 18080 Granada, Spain \\
   $^8$ Donostia International Physics Centre, Paseo Manuel de Lardizabal 4, 20018 Donostia-San Sebastian, Spain \\ $^9$ Ikerbasque, Basque Foundation for Science, 48013 Bilbao, Spain \\
   $^{10}$ Instituto de Física, Universidade Federal da Bahia, 40210-340,
Salvador, BA, Brazil \\
    $^{11}$  Department of Astronomy, University of Michigan, 311 West Hall,
1085 South University Ave., Ann Arbor, USA \\
    $^{12}$ Departamento de Astronomia, Instituto de Astronomia, Geofísica
e Ciências Atmosféricas da USP, Cidade Universitária, 05508-900,
São Paulo, SP, Brazil \\
$^{13}$ Instruments4, 4121 Pembury Place, La Cañada-Flintridge, CA 91011,
USA \\
$^{14} $ Departamento de Física,
Universidade Federal do Espírito Santo, 29075-910, Vitória, ES,
Brazil \\
$^{15}$  INAF – Osservatorio Astronomico di Trieste, via Tiepolo 11, 34131
Trieste, Italy \\
$^{16}$ IFPU – Institute for Fundamental Physics of the Universe, via
Beirut 2, 34151 Trieste, Italy \\
$^{17}$ Departamento de Astrofísica, Universidad de La Laguna, 38206,
La Laguna, Tenerife, Spain \\
$^{18}$ Instituto de Astrofísica de Canarias, 38200 La Laguna, Tenerife,
Spain             }

   \date{}

  \abstract
    {}
    {We investigated the impact of photometric redshift errors in the ongoing Javalambre Physics of the Accelerating Universe Astrophysical Survey (J-PAS) on void identification and properties using a watershed-based method, aiming to assess the recovery of individual voids and the overall void environment.}   
    {We created galaxy mock catalogues for redshift z = 0.1 using the IllustrisTNG300-1 simulation, defining two datasets: an \textit{ideal} sample ($m_r < 21$ mag) and a \textit{perturbed} sample with the Z-coordinate errors mimicking J-PAS's line-of-sight errors, derived from the precursor miniJPAS survey data. We identified voids using ZOBOV, a watershed algorithm.}
    {We found 1065 voids in the \textit{ideal} sample and 2558 voids in the \textit{perturbed} sample. The \textit{perturbed} sample voids have, on average, smaller sizes and denser interiors. We filtered out voids based on density and radius in order to eliminate overdense and small spurious instances. The stacked density profile of filtered voids in the \textit{perturbed} sample remains close to the average density even at the boundary peak, indicating a strong blurring of structures by the redshift errors. The number of \textit{ideal} sample voids for which at least $50\%$ of the volume is recovered by a void in the \textit{perturbed} sample is 53 (29 for the filtered sample). The volume occupied by these voids is less than $10\%$ of the simulation volume.  Merging voids in the \textit{perturbed} sample marginally improves the recovery.
    The overall volumes defined as voids in the two samples have an overlap of $80\%$, making up $61\%$ of the simulation box volume.}
    {While some statistical properties of voids might be recovered sufficiently well, the watershed algorithms may not be optimal for recovering the large-scale structure voids if applied straight to photometric redshift survey data.}
    {}


   \maketitle

\section{Introduction}

Cosmic voids are underdense regions embedded in a network of overdense filaments and superclusters in the large-scale structure of the Universe. They occupy the majority of the volume in the Universe, are important cosmological probes, and are an indispensable environment for studying star formation processes and galaxy assembly. This makes them an essential research target both for cosmology and galaxy physics.

Some of the early observations of large voids in galaxy spatial distribution were made by \cite{1978MNRAS.185..357J, 1987ApJ...314..493K} and were later confirmed in the first galaxy redshift surveys such as the Center for Astrophysics survey \citep[CfA;][]{1986ApJ...302L...1D}. 
A few decades later, the galaxy distribution maps provided by the 2dFGRS and SDSS surveys cemented the idea that voids are an essential and prominent component of the cosmic web \citep[see e.g.][]{colless20032df, Hoyle_2004, Pan_2012}. 

The main theoretical models for void formation are spherical/ellipsoidal evolution models \citep[see e.g][]{1982ApJ...262L..23H, 1984MNRAS.206P...1I, 1985ApJS...58....1B, 1992ApJ...388..234B, 2004MNRAS.350..517S} which state that a void is a region of space that contains around $20\%$ of the average matter density in the Universe, and that presents steep, overdense boundaries. More concretely, voids present a reverse top-hat density profile, with the number density of galaxies rising very rapidly as one departs from the void centre, reaching an overdense peak and then slowly decreasing towards the average \citep{2004MNRAS.350..517S}.

Voids formed as a consequence of tiny initial perturbations in the primordial Gaussian density field. According to the gravitational instability theory, these fluctuations displaced matter, which gave rise to overdense regions that over time experienced an inward gravitational force, causing them to attract more of the surrounding matter and thus increase in density \citep{1998ARA&A..36..599B}. The primordial overdensities evolved in a bottom-up, hierarchical manner followed by decoupling from the Hubble flow and a gravitational collapse which led to the formation of clusters, filaments, and walls as predicted by the Cold Dark Matter model \citep[see e.g][]{1974ApJ...187..425P,1991ApJ...379..440B,1993MNRAS.262..627L,LIDDLE19931}.  As these regions attract matter, they leave behind underdense gaps: the cosmic voids. The reverse also applies, which means that voids experience a repulsive gravity effect 
as their expansion rate starts exceeding the Hubble flow \citep[raising the question of how to calculate the average expansion, see e.g][]{2000PhRvD..62d3525B, 2004JCAP...02..003R, 2018A&A...610A..51R, 2023arXiv231019451V, 2024arXiv240104293C}.
For this reason, they are often referred to as super-Hubble bubbles, pushing the existing matter within them toward their boundaries and forming a ridge. This evacuation of matter will leave the void increasingly empty as time goes by. In the non-linear stages, a void experiences shellcrossing: shells of matter initially located inside will overtake shells located at the void ridge \citep{1985ApJS...58....1B}.

Voids also present hierarchical levels:
small voids embedded in a large underdensity tend to expand with time, squeeze, and evacuate the matter located between them. At non-linearity, the small voids will tend to collide and merge, dissolving any remaining structure and giving birth to a single, large, all-encompassing void (also known as the void-in-void process). On the other hand, a void can also be surrounded by large overdensities (void-in-cloud), which will tend to collapse and squeeze the void out of existence, essentially restricting the void size function \citep{1993MNRAS.263..481V, 2004MNRAS.350..517S}.

Despite their low matter content, voids represent a fundamental feature of the cosmic web. For example, they are the most voluminous component of the web, with occupation volumes between $70\%$--$90\%$ \citep{Ganeshaiah_Veena_2019, Hellwing_2021, 2024ApJ...962...58C}.  They have been used to estimate cosmological parameters \citep{2010MNRAS.403.1392L, 2014MNRAS.443.2983S, Hamaus_2016,  2024A&A...682A..20C}, measure the baryonic acoustic oscillations \citep{Kitaura_2016, 2023MNRAS.526.2889T} and restrict the neutrino mass \citep{Schuster_2019, 10.1093/mnras/stz1944} among others. For a review of the importance of voids in cosmology, e.g. see, \cite{2019BAAS...51c..40P, bromley2024cosmologyvoids}.

Furthermore, the low-density environment that voids offer is an ideal laboratory for studying the formation and evolution of galaxies. Since voids are mostly devoid of matter, galaxies located within them will be at a more primitive stage of evolution due to the lack of interactions with other galaxies. As such, one can study the evolution of galaxies independently without having to take into account events such as galaxy merging, which is far more prevalent in the denser regions of filaments \citep{2015A&A...576L...5T} or superclusters \citep{2020A&A...641A.172E, 2021A&A...649A..51E}. A morphology-density relation has been known since \cite{1980ApJ...236..351D}, which states that elliptical galaxies populate higher-density central regions of clusters while spiral galaxies reside in sparser regions. Since then, void galaxies have been studied extensively in both simulations and observations and were shown to present properties different from galaxies located in denser environments. These include fainter and bluer appearances, lower stellar masses, higher star formation rates and higher spin parameters \citep{Grogin_1999, Kreckel_2012, 2012MNRAS.426.3041H, 10.1093/mnras/stw2362, Rosas_Guevara_2022,2022A&A...668A..69E, 2024ApJ...962...58C, 2024A&A...687A..98C} \citep{2021MNRAS.505.1223P, 2024MNRAS.527.4087J}.

Despite current advancements in our understanding of voids, there is still no universally accepted definition of what a void is. As a result, various void-finding algorithms have been developed, such as spherical void finders \citep{2002ApJ...566..641H, 2005MNRAS.360..216C, 2005MNRAS.363..977P}, watershed based-methods \citep{2007MNRAS.380..551P, 2008MNRAS.386.2101N, 2014ascl.soft07014S}, hybrid methods \cite{Jennings_2013}, 2D projections \cite{2015MNRAS.454.3357C}. For a comparison of how well various void algorithms manage to identify voids, we refer to \cite{2008MNRAS.387..933C}.

So far, void-finders have been primarily applied to galaxy spectroscopic redshift catalogues, as photometric redshifts are less accurate and introduce smearing in the line-of-sight (LOS) positions of galaxies. One attempt to take this issue into account was made by \cite{2017MNRAS.465..746S}, who projected galaxies into 2D slices and identified voids in the 2D density field of the slice. They found that voids are recovered best from a spectroscopic sample in an equivalent photometric one for redshift slices with a thickness comparable with the photo-$z$ error. Furthermore, they found that larger voids are less affected by the smearing of galaxies due to the redshift errors. At the time of writing, there are no other studies that focus on the effect that photometric redshift errors have on void identification and properties.

This paper aims to explore how the uncertainty in position along the line of sight, caused by redshift errors, impacts the identification and features of voids using a watershed void-finding algorithm in the ongoing Javalambre Physics of the Accelerating Universe Astrophysical Survey \citep[J-PAS,][]{jpas_article, 2021A&A...653A..31B}. In particular, we are interested in two aspects: (1) how well is the void population and its properties recovered in general and (2) how well is the void environment recovered for each specific location in space. While (1) is important for cosmological applications, (2) is especially interesting from the point of view of galaxy evolution.

At the time of writing, J-PAS has not yet observed a large enough volume that could allow the study of voids. Thus, to overcome this issue, we generate realistic galaxy mocks based on the IllustrisTNG300-1 hydrodynamic simulation.

This paper is organised as follows. In Section~\ref{sec:dat_and_meth}, we describe the J-PAS survey and the galaxy mocks based on IllustrisTNG300-1, the implementation of redshift errors, and the watershed-based void identification algorithm used in this work. In Section~\ref{sec:results}, we present our results: the properties of voids (such as void abundance and radii, density parameters, and stacked profiles) obtained within galaxy mocks and make relevant comparisons with the literature. Subsequently, we present the recovery of individual voids and of the overall void environment. Finally, in Section~\ref{sec:disc_and_conc}, we summarise our work and discuss limitations and comparisons with other studies.

\section{Data and \textbf{methods}}
\label{sec:dat_and_meth}

Cosmic voids often span up to a few tens of megaparsecs. Since J-PAS has not yet observed a sufficiently large contiguous area of the sky in order to permit the study of such large structures, we used the IllustrisTNG300-1 simulation to construct galaxy mock catalogues. To maximally mimic the expected J-PAS data, we took into account J-PAS magnitude limits and considered realistic photometric redshift errors from miniJPAS obtained through TOPz \citep{2022A&A...668A...8L}.

\subsection{The J-PAS survey}
The J-PAS survey \citep{jpas_article, 2021A&A...653A..31B} is conducted at the Obervatorio Astrofisico de Javalambre (OAJ), developed by the Centro de Estudios de Física del Cosmos de Aragón (CEFCA).
J-PAS uses the Javalambre Survey Telescope (JST/T250), a Ritchey-Chretien telescope with an aperture of 2.55~m and 3 degree FoV. J-PAS stands out for its innovative filter system, comprising 54 narrow band filters spanning a wavelength range of 3780 {\AA} to 9100 {\AA}, along with 2 intermediate band filters and one broad filter.
A continuous spectrum coverage is created by the narrow band filters due to their FWHM of 145 {\AA} and the 100 {\AA} spacing in between. The survey will cover an area of 8500 deg$^{2}$ of the Northern Sky and it is estimated to measure the photometric redshifts of 9 $\times$ 10$^{7}$ Luminous Red Galaxies and Emission Line Galaxies, several million Quasi-Stellar Objects and $7\times 10^{5}$ galaxy clusters and groups up to a $z = 1.3$.

One method to estimate the photometric redshifts in J-PAS is the TOPz workflow \citep{2022A&A...668A...8L}, which provides template-based photo-$z$ estimation with added J-PAS-specific features. So far, TOPz has provided accurate redshifts for the miniJPAS survey, which fulfilled the expectation for the J-PAS: $38.6\%$ of galaxies with $m_{\rm r} < 22$ mag have reached the target goal of a photometric redshift error $dz/(1+z) < 0.003$. Further improvements over miniJPAS are expected as J-PAS will release more data.

\subsection{J-PAS mocks with IllustrisTNG}

IllustrisTNG \citep{2018MNRAS.480.5113M,2018MNRAS.475..676S,2018MNRAS.475..624N, 2018MNRAS.477.1206N,2018MNRAS.475..648P} represents the successor of the original Illustris project \citep{2014MNRAS.445..175G} -- a set of cosmological hydrodynamic simulations aimed to study the galaxy formation and evolution. The IllustrisTNG300 includes the addition of magnetic fields, improvement of galactic winds, and AGN feedback. The cosmological parameters are taken from the Planck 2015 results: $\Omega_\Lambda = 0.6911$, $\Omega_m = 0.3089$, $b = 0.0456$, $\sigma_8 = 0.8159$, $n_s = 0.9667$, and $H_0 = 0.6774$~km$\,$s$^{-1}\,$Mpc$^{-1}$ \citep{2016A&A...594A..13P}. The simulation uses 2500$^3$ dark matter particles and 2500$^3$ gas particles in a comoving box with a side length of 205\mhp. The mass resolutions of dark matter and gas, respectively, are $5.9 \times 10^7 M_{\odot}$ and $1.1 \times 10^7 h^{-1} M_{\odot}$. We decided to work with the IllustrisTNG300-1 model since it was designed to model the evolution of galaxies simultaneously with the large-scale structure and to provide us with a sufficiently large galaxy catalogue.

In the TNG simulation, a galaxy is defined as "any luminous (sub)halo, i.e. any gravitationally bound object with a non-zero stellar component" \citep{2018MNRAS.475..648P}. We apply a magnitude limit, corresponding to our miniJPAS sample (see below) to form our TNG galaxy sample. This criterion guarantees that the galaxies have realistic properties.

In order to mimic the J-PAS galaxy data with the IllustrisTNG simulation, we needed to address the J-PAS redshift errors, which potentially are the main limiting factor for the identification of voids. To this end, we turned to the miniJPAS \citep{2021A&A...653A..31B}, the precursor survey of J-PAS carried out between May and September of 2018. The survey used the main J-PAS telescope JST/T250 and covered a field of $\approx$~1~deg$^{2}$. The observations were made through 54  narrow-band filters and six medium and wide-band filters using the JPAS-Pathfinder camera. The miniJPAS dataset covered by the DEEP2 and DEEP3 spectroscopic surveys comprises of 4457 galaxies with $m_{\rm r}$~<~23~mag up to a $z$~=~1.5 and has been used to study, among other topics, the stellar populations of nearby galaxies, and the large-scale structure including detecting groups and clusters \citep[see e.g][]{2021A&A...649A..79G, Maturi_2023}.

\subsection{Redshift error modelling}

We used the miniJPAS redshift errors as provided by the TOPz workflow \citep{2022A&A...668A...8L}.
TOPz employs a template-based photo-$z$ estimation tweaked to take advantage of J-PAS-specific features.
Based on the miniJPAS observations, spectral templates of galaxies are created using the synthetic galaxy spectrum generation software CIGALE \citep{2019A&A...622A.103B}. 
To assess the accuracy of the redshift estimation, \cite{2022A&A...668A...8L} used spectroscopic redshifts from the DEEP2, DEEP3, and SDSS surveys, available for 1989 miniJPAS galaxies with $m_{\rm r}$~<~22. The redshift errors were calculated as
    
\begin{align}\hspace{0.3cm}
    z_{\rm err} = \frac{|z_{\rm phot} - z_{\rm spec}|}{1 + z_{\rm spec}},
\end{align}
where $z_{\rm phot}$ is the estimated photometric redshift while $z_{\rm spec}$ is the spectroscopic redshift, used as the ground truth.

While the photometric redshift errors of (mini)J-PAS are already relatively small, \citep{2021A&A...654A.101H, 2022A&A...668A...8L, 2024A&A...684A..61H}, the accuracy of photometric redshifts estimates can be further improved by selecting a subsample based on the \textit{odds} parameter, which correlates well with $z_{err}$ \citep{2021A&A...654A.101H}. This parameter gives the relative area of the estimated redshift PDF within some user-defined interval centred on the estimated redshift. An \textit{odds} value closer to one indicates that the distribution is narrow and centres around the highest PDF value, whereas low values of \textit{odds} indicate a broad area, where any given redshift estimate would have a low probability.

We chose to mimic the case of a nearby sample of J-PAS galaxies using the IllustrisTNG galaxy catalogue at a redshift snapshot $z = 0.1$. At this redshift value, an error of $0.003(1+z)$ corresponds to a comoving radial distance error of $\approx$ 14.5~Mpc. This already suggests a challenge in identifying void galaxies in J-PAS.
For ascribing realistic J-PAS redshift errors to these IllustrisTNG galaxies, we first defined a miniJPAS local reference sample, which we call "No \textit{odds} cut", containing 156 galaxies at $z_{\rm spec} < 0.15$. We defined a second reference sample ("\textit{odds} cut") by restricting the \textit{odds} parameter to values higher than 0.99, which reduces the sample to 79 galaxies. The outlier fraction (i.e. the galaxies with a photometric redshift error worse than 0.05) is $9\%$ for the "No \textit{odds} cut" sample and $0\%$ for the "\textit{odds} cut" sample.

\begin{figure}[t]
    \includegraphics{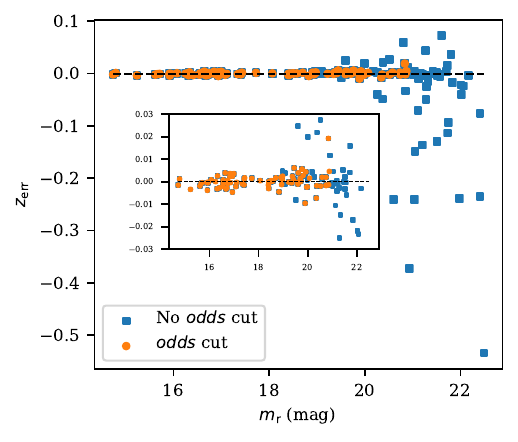}
    \caption{Dependence of miniJPAS redshift errors on apparent magnitude. The "No \textit{odds} cut" sample (blue, squares) is obtained by imposing a redshift cut in MiniJPAS at $z_{\rm spec} < 0.15$, while the "\textit{odds} cut" sample (orange, circles) has an additional cut in the \textit{odds} value parameter (see text for details). The inset shows a zoom-in region centred on zero for a better visual inspection.}
    \label{minijpaserr}
\end{figure}

In Fig. \ref{minijpaserr}, we show the correlation between photometric redshift errors and the $r$-band magnitude for these two miniJPAS samples. Since the \textit{odds} cut effectively translates into a magnitude cut $m_{\rm r} < 21$, we use the latter limit to define our IllustrisTNG galaxy catalogue, for which the \textit{odds} parameter is not applicable. The resulting sample contains 358~186 galaxies within a periodic box with a volume of 8.6 $\times$ 10$^6$ (Mpc $h^{-1}$)$^3$. Fig. \ref{minijpaserr} also shows that the applied \textit{odds} limit removes galaxies with the worst redshift errors without a too heavy toll on the sample size.

Before applying the miniJPAS redshift errors to the IllustrisTNG galaxies, we first translated these redshift errors into distance errors, accounting for the IllustrisTNG snapshot being situated at $z = 0.1$. To do this, we added the miniJPAS redshift errors to the redshift of the IllustrisTNG snapshot ($z_{0.1}$) as expressed in Eq.~\ref{zpert}. The resulting perturbed redshifts ($z_{\rm pert}$) were then converted into corresponding perturbed luminosity distances ($d_{\rm pert}$)\footnote{Astropy \citep{astropy:2013, astropy:2018, astropy:2022}}. Finally, the distance errors ($d_{\rm err}$) were calculated by subtracting the perturbed distances ($d_{\rm pert}$) from the fixed luminosity distance at $z = 0.1$ ($d = 475$~Mpc).

\begin{align}\hspace{0.3cm}
    \label{zpert}
    (1 + z_{\rm 0.1}) (1 + \pm z_{\rm err}) &= (1 + z_{\rm pert}), \\
    z_{\rm pert} &\rightarrow d_{\rm pert}, \\
    d_{\rm err} &= d_{z=0.1} - d_{\rm pert}.
\end{align}


Since Fig. \ref{minijpaserr} indicates that there is a substantial increase in the error amplitude beyond $m_{\rm r} \approx 19$  (standard deviations of 0.002 for $m_{\rm r}$ < 19 and 0.004 for $m_{\rm r}$ > 19, respectively) we split the miniJPAS sample into two magnitude bins accordingly. We did the same magnitude separation for IllustrisTNG galaxies by first computing their apparent magnitudes considering the snapshot redshift $z_{0.1}$ and their absolute $r$ magnitudes.
We then randomly sampled $d_{\rm err}$ considering the two magnitude bins and added them to the $Z$-coordinate of galaxies\footnote{As previously noted, the distance errors obtained are actually \textit{luminosity} distance errors. In principle, it would have been correct to use comoving distance errors instead. The current errors are slightly overestimated (about 10$\%$), but this does not significantly impact the overall results.}. The resulting standard deviation values for the distance errors are 7.1~Mpc for $m_{\rm r}$ < 19 and 16.5~Mpc for $m_{\rm r} > 19$.

As a result, we obtained two galaxy mock samples: the \textit{ideal} J-PAS (IllustrisTNG galaxies with $m_{\rm r} < 21$) and the \textit{perturbed} J-PAS, where we additionally perturb the coordinates along one axis to mimic the miniJPAS redshift errors for such galaxies. In the upper panel of Fig.~\ref{voidctrimp}, we show plots of the \textit{ideal} and \textit{perturbed} galaxy mock samples in two different projections of the box. The smearing of galaxies due to the redshift errors is especially visible in the top right $X-Z$ panel of the second plot.

We note that the redshift and odds cut that we impose on miniJPAS result in a very low number of galaxies (79), which we use to ascribe errors to a large number of galaxies (358~186). This approach may introduce artefacts in the galaxy distribution. We also tried the inverse transform sampling of the error distribution using both linear interpolation and the Gaussian approximation. The results from linear interpolation were quantitatively similar to the ones presented below, whereas the Gaussian approximation significantly overestimated the errors.

\subsection{Watershed voids}
\label{sec:vide}
Once we had defined our galaxy mocks, we proceeded to the void identification with ZOnes Bordering On Voidness \citep[ZOBOV][]{2008MNRAS.386.2101N}. For this task, we used VIDE \citep{2014ascl.soft07014S}, an often-employed pipeline of the popular watershed transform method  \citep{2007MNRAS.380..551P} used to detect voids in redshift surveys and N-body simulations. The pipeline is built on the ZOBOV algorithm, which computes the Voronoi tessellation of the tracer distribution in order to estimate the density field. The tessellation provides us with the minimum density values in our sample, the seeds of our voids. Subsequently, the watershed transform is applied in order to group the Voronoi cells together into voids. This allows us to reconstruct the void population and to preserve the shapes of voids (without any prior assumption of their geometry).

The pipeline internally computes various standard void properties. For example, it provides the void macrocentre defined as:

\begin{align}\hspace{0.3cm}
    X_{\rm v} = \frac{1}{\sum V_{\rm i}} \sum x_{\rm i} V_{\rm i}, 
    \label{eqcenter}
\end{align}
where $x_i$ and $V_{\rm i}$ represent the positions and Voronoi volumes of the galaxy tracers, respectively. VIDE also provides the equivalent radius of a void, defined as the radius of a sphere with the same volume as the void:

\begin{align}\hspace{0.3cm}
    R_{\rm eq} = \left(\frac{3}{4 \pi} V \right)^{\frac{1}{3}}
    \label{eqrad},
\end{align}
where $V$ is the volume of the void, given by the sum of all the individual Voronoi cells. Furthermore, it offers easy access to the void member galaxies and allows one to study their properties and evolution. ZOBOV has been previously used in both observations \citep{2019MNRAS.490.3573F, 2020JCAP...06..012H, 2022AAS...24013923D} and simulations \citep{2015JCAP...11..036H, 2022A&A...667A.162C,  2022ApJ...935..100K}, and thus it will allow us a fair comparison with existing results.
Taken together, we considered this watershed-based method to be a suitable void identification method for our research goals of the J-PAS project.

\begin{figure}[t]
    \centering
    \includegraphics[width = 9cm]{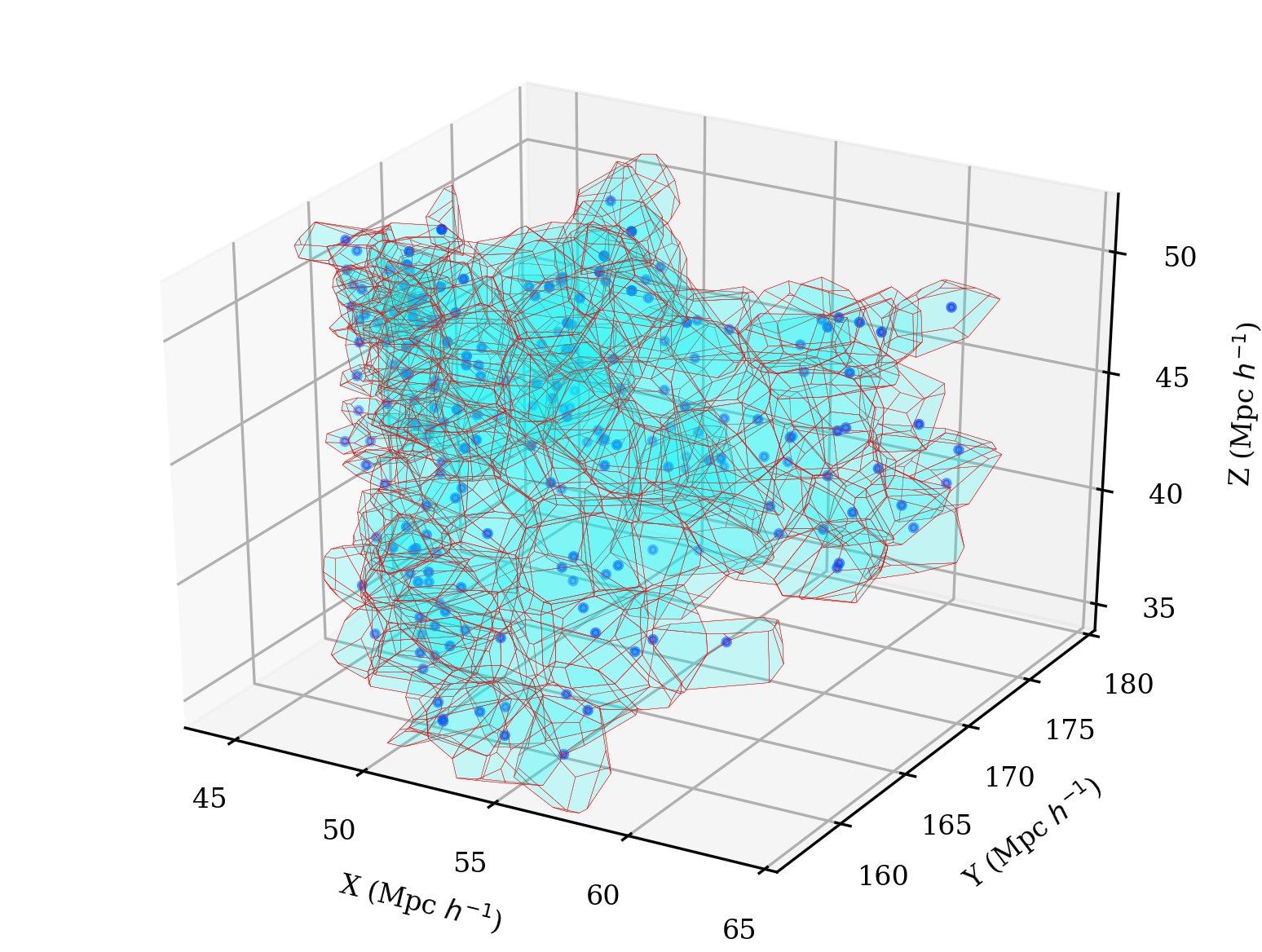}
    \caption{Example of a void identified by ZOBOV. The void volume consists of Voronoi cells (red and cyan polygons) corresponding to its member galaxies (blue dots). The void contains 202 galaxies (and correspondingly consists of the same number of Voronoi cells) and encompasses a volume of about 2700~(Mpc$/h)^3$. The minimum void density $\rho_{\rm min}$ = 0.30 $\bar{\rho}$.}
    \label{voidex}
\end{figure}

We ran the VIDE pipeline on our J-PAS mock galaxy catalogue in order to obtain the void catalogues. We show an example of a detected void and its corresponding Voronoi cells in Fig.~\ref{voidex}.

Since ZOBOV is designed to find local minima in a density field and merge Voronoi cells until they form voids, some of the identified voids would actually be spurious. For example, it has already been shown that the ZOBOV algorithm manages to identify voids in a Poisson distribution \citep{2008MNRAS.386.2101N}. As a consequence, these spurious voids will present properties -- such as high densities -- which are in disagreement with the theoretical models of void formation \citep{2004MNRAS.350..517S, 1993MNRAS.263..481V}. Indeed, \cite{2014MNRAS.440.1248N} have shown how ZOBOV can give rise to overdense voids and cautioned that density criteria should be considered before using voids in cosmological tests.

As a result, we need to define specific criteria in order to clean the void catalogue of spurious detections. To establish these criteria, we first need to define some quantities. The void density $\rho_{\rm void}$ represents the ratio between the number of galaxies that define the void and the void volume. The minimum void density, $\rho_{\rm min}$, represents the lowest density of an associated Voronoi cell of a galaxy in a void. Note that this does not necessarily represent the density value at the void centre since voids usually do not contain matter there. Instead, this gives the density value associated with the largest volume of a Voronoi cell that comprises a void.  We describe how we imposed criteria on these parameters in Sect.~\ref{sec:results}.

Furthermore, VIDE allows the user to merge voids and create a nested void hierarchy by modifying the initial density threshold parameter. This is an internal parameter of VIDE that is tweaked before the ZOBOV algorithm is applied to the galaxy distribution. This parameter essentially tells ZOBOV which initial voids should be merged together based on their density according to a threshold value. The parameter is initially set to a very low value, normalised to the average density ($\rho_{\rm threshold} = 1 \times  10^{-9}$), such that ZOBOV does not merge any voids and this is the value that has been used in our initial analysis. However, one may tweak this value and create a void hierarchy: zones with densities lower than the threshold are grouped together. A $\rho_{\rm threshold} = 1$ will essentially merge all the voids together into a single, parent supervoid that spans the whole volume of the box and which contains all the other children (sub-voids) grouped in different hierarchical levels. Intermediate values will give rise to various parent voids, potentially containing many children that occupy separate sub-volumes within the parent, separated by a ridge line. Voids that do not have any children are also a possibility. We discuss different merging thresholds used in this paper in Section 3.2.1.

In order to assess the voids in 3D space, for example for calculating the overlap of voids between different data sets, we mapped the detected voids to a grid using the discrete Voronoi tesselation.

\section{Results}
\label{sec:results}

In this paper, we addressed the following questions: \textit{How} do J-PAS-like photometric redshift errors of galaxies affect the general properties of the identified void population (i.e., the number of detected voids, void radii, and void density profiles)? \textit{How} well can we recover individual "true" voids (identified from a galaxy mock dataset without redshift errors) from a dataset in which we model redshift errors? \textit{How }well can we recover the overall void environment (e.g. at each volume element)?

In order to answer these questions, we created two J-PAS galaxy mock datasets using the IllustrisTNG simulation: the \textit{ideal} J-PAS sample, where we assumed no photometric redshift errors, and the \textit{perturbed} J-PAS sample, where we perturbed the $Z$-coordinate of galaxies with some distance errors corresponding to the actual errors occurring in miniJPAS, the precursor survey of J-PAS.

In this section, we present the results of the void identification obtained with the watershed ZOBOV algorithm. 
We show various properties of voids such as density parameters (Fig.~\ref{impdensparam_filt}),  cumulative volume occupation distribution (Fig.~\ref{volcdf}), void abundance and radii (Fig.~\ref{radiiimp}), and stacked radial density profiles (Fig.~\ref{combined_dens_prof}) presenting comparisons between voids found in the \textit{ideal} and \textit{perturbed} samples. 
We implemented various density- and radius-based filtering criteria to increase the reliability of the detected voids (Table~\ref{criteria_stats}). 

One of the important tasks in our work is to recover voids in the \textit{ideal} sample using the data from the \textit{perturbed} sample.
We looked at how well we can recover individual voids, setting $50\%$ volume overlap as the threshold. Subsequently, we also studied how much of the overall void environment can be recovered (i.e. the total volume covered by voids in the two samples).

\subsection{Void properties in J-PAS mocks}

\begin{figure*}[t]
    \centering
    \begin{minipage}[t]{0.49\textwidth}
        \centering
        \includegraphics[width=\textwidth]{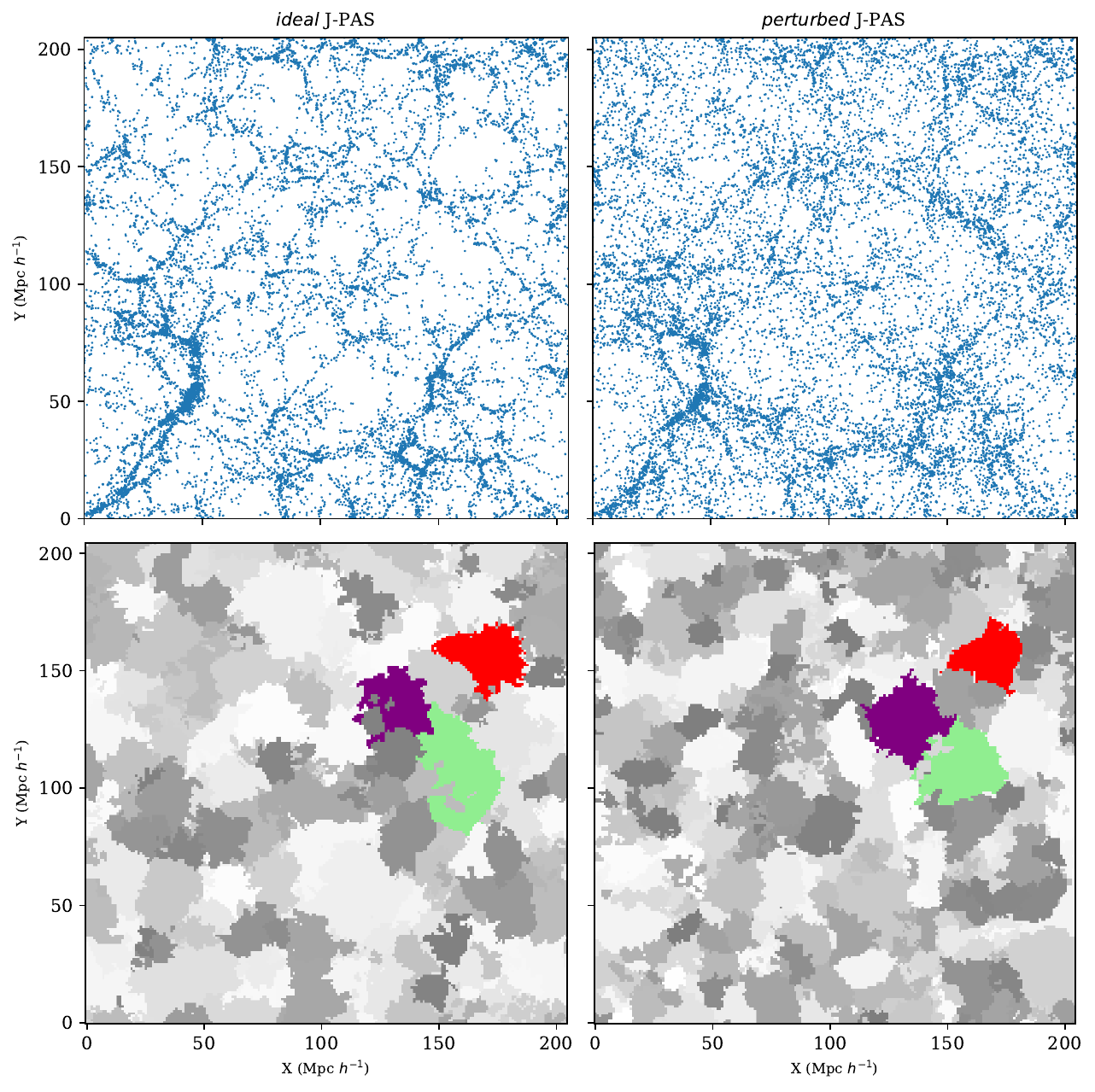}
        \caption*{}
    \end{minipage}
    \hfill
    \begin{minipage}[t]{0.49\textwidth}
        \centering
        \includegraphics[width=\textwidth]{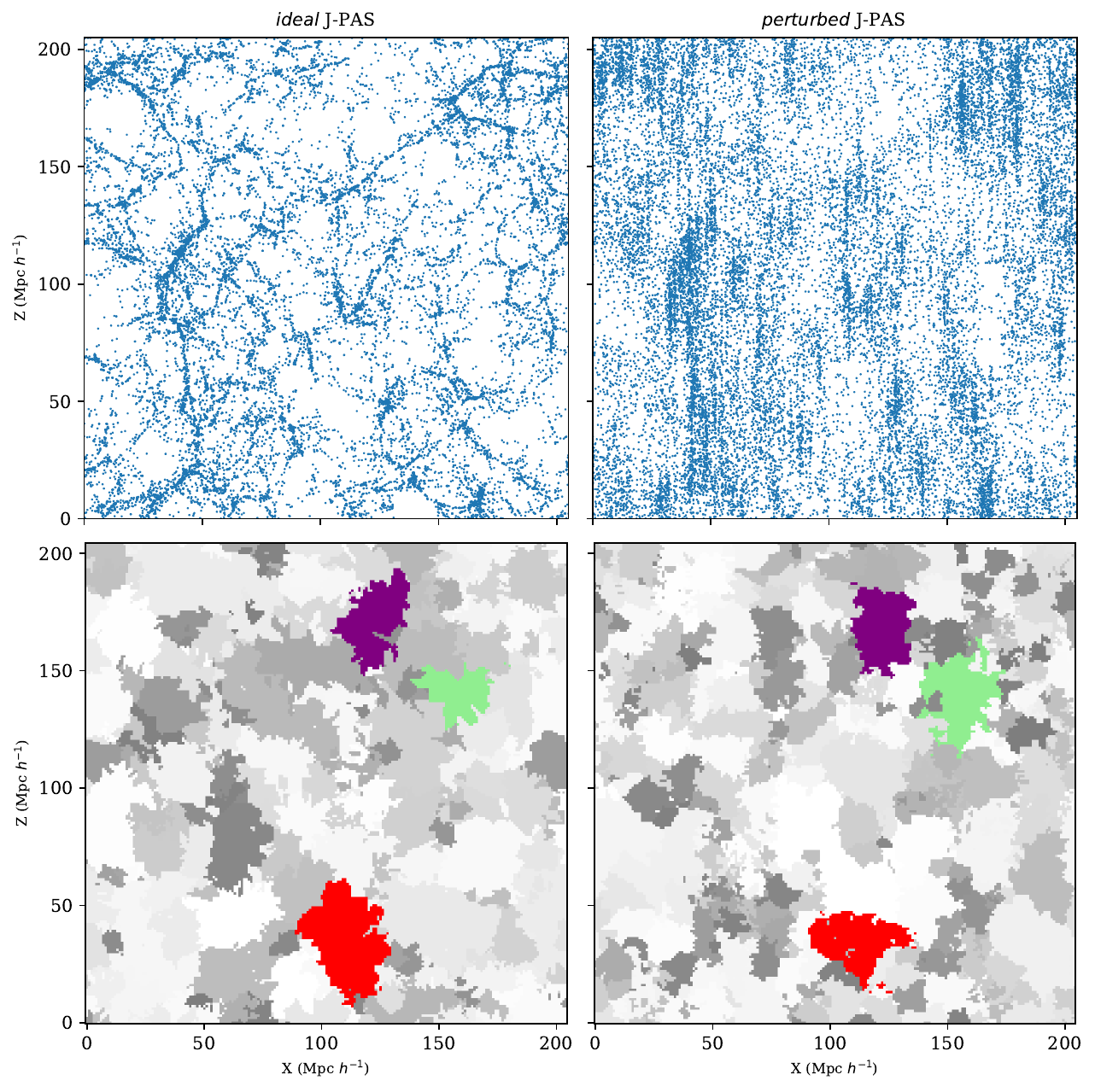}
        \caption*{}
    \end{minipage}   
    \caption{\textit{Ideal} and \textit{perturbed} J-PAS galaxy mock samples (top panels) and void distributions (bottom panels) in $X$–$Y$ and $X$–$Z$ projections. We show a 10 Mpc-thick slice for each projection. The effect of redshift errors on the galaxy distribution is evident in the $X$–$Z$ projection of the \textit{perturbed} sample (fourth column, top panel). Cross-sections of voids are shown in greyscale, while colours highlight voids that have more than 50\% volume overlap between the \textit{ideal} and \textit{perturbed} samples (see Sec. \ref{void_ov} for further discussion). A visual comparison reveals that voids detected in the \textit{ideal} sample tend to fragment into various-sized voids in the \textit{perturbed} sample.}
    \label{voidctrimp}
\end{figure*}

Using the ZOBOV algorithm, we identified 1065 voids in the \textit{ideal} sample and 2558 voids in the \textit{perturbed} sample of our mock J-PAS datasets. In Fig.~\ref{voidctrimp}, we show the galaxy distributions and the cross-sections of voids in a 10~Mpc-thick slice of the \textit{ideal} and \textit{perturbed} samples in two different projections of the box. The familiar elements of the cosmic web, such as filaments, superclusters, and voids, can be easily recognised in the $X$--$Y$ projections of the \textit{ideal} and \textit{perturbed} samples.
Due to the watershed transform retaining the topology of the field, one can immediately notice that voids are not isotropic structures but instead come in a variety of shapes and sizes.
One can also already see how the displacement of galaxies due to redshift errors impacts the galaxy distribution: a shift in the $Z$-direction moves galaxies in and out of the slice.
Most notably, the underdense areas of the \textit{perturbed} sample seem denser in the $X$--$Y$ plane. 
It is possible to note (although less clearly) that structures like clusters and filaments also become more diluted.
As expected, the effect of redshift errors is especially visible in the $X$--$Z$ projection (top right panel), where galaxies appear smeared across the figure vertically.

A visual inspection of the general population of voids (Fig. \ref{voidctrimp}, bottom panels, shown in grey) in the \textit{ideal} and \textit{perturbed} samples confirms that there are more \textit{perturbed} sample voids than \textit{ideal} sample ones. 
This is most likely explained by voids in the \textit{ideal} sample, which get fragmented into voids of various sizes in the \textit{perturbed} sample (see also the right panel of Fig. \ref{3dvol}). This comes as a direct consequence of the distance errors that shift the galaxy positions, which causes the ZOBOV algorithm to segment the galaxy population into different Voronoi cells that will group together into different voids. Thus, the void identification is highly sensitive to the precise position of galaxy tracers. Despite this fragmentation, one can see how some of the voids in the ideal and perturbed sample still correlate (we show some of these voids with colours).

\subsubsection{Void density parameters}

\begin{figure}[t]
    \centering
    \includegraphics{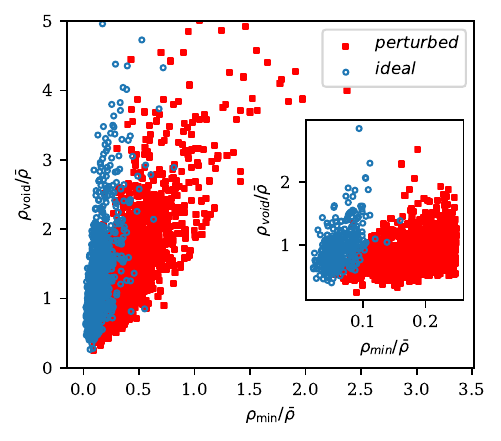}
    \caption{\textit{Main plot}: Void density parameters. The plot shows the correlation between the overall void density and the minimum void density for the unfiltered \textit{ideal} (blue open circles) and  \textit{perturbed} samples (red squares). \textit{Inset}: same for medium-filtered voids. Both density parameters have been normalised to the average density in the box, $\bar{\rho}$.}
    \label{impdensparam_filt}
\end{figure}

It is natural to assume that voids are regions with densities generally below the Universe's average density. However, the ZOBOV algorithm segments the whole galaxy population into voids; thus, spurious voids or voids within dense regions (e.g. superclusters), with average densities above the average of the Universe, should also be expected. We examined the situation by looking at the density parameters of the detected voids. Following the method of \cite{2014MNRAS.440.1248N}, we plot the correlation between the void density $\rho_{\rm void}$ and the minimum void density, $\rho_{\rm min}$, for the voids in the \textit{ideal} and \textit{perturbed} samples (Fig.~\ref{impdensparam_filt}). 
Both parameters were normalised in units of the average density over the simulation volume $\bar{\rho}$.

We found that 631 voids (out of 1065, i.e. 59\%) in the \textit{ideal} and 1644 voids (out of 2558, i.e. 64\%) in the \textit{perturbed} samples have void densities larger than the average galaxy density. 
On average, both void samples present $\bar{\rho}_{\rm void}$ $\approx$ 1.3 $\bar{\rho}$. 
However, the voids in the \textit{ideal} sample have a lower average minimum density ($\bar{\rho}_{\rm min}$ $\approx$ 0.13 $\bar{\rho}$) than the \textit{perturbed} sample voids ($\bar{\rho}_{\rm min}$ $\approx$ 0.36 $\bar{\rho}$) indicating that their interior regions are emptier.

A commonly used definition of a void is based on the spherical model \citep[see e.g][]{1985ApJS...58....1B, 2004MNRAS.350..517S}, which states that a void is fully formed when it reaches an average density of $\approx 0.2 \bar{\rho}$ at the shell-crossing stage. Figure~\ref{impdensparam_filt} shows that a substantial number of voids in our galaxy mocks have values higher than that. This is again the result of ZOBOV dividing up the whole volume of the simulation box into voids and thus assigning overdense filaments and walls to voids as their boundaries. This, in turn, raises the average density within voids. As such, obtaining voids by imposing a criterion on $\rho_{\rm void}$ in voids can lead to erroneous results. Instead, we decided to impose cuts based on the $\rho_{\rm min}$ value. In this way, we ensure that the starting seeds from which the algorithm builds up voids are underdense and any spurious overdense regions will thus be eliminated. 

In addition to restricting the minimum void density, the total volume that voids occupy in the simulation box has to be taken into account as well. Previous work has shown that the void volume filling fraction in simulations ranges between $70\%$--$90\%$ \citep{Ganeshaiah_Veena_2019, Hellwing_2021,2024ApJ...962...58C}.

Taking these conditions into account, we defined three different levels of filtering for the voids in the \textit{ideal} and \textit{perturbed} samples: soft, medium, and strong. See Table~\ref{criteria_stats} for a summary of the way these filters were defined and of the resultant void statistics. We focus on the medium level for the rest of the analysis of void properties and provide some comparisons with the soft- and strong-filtered void samples at the end.

\begin{table*}[t]
\captionsetup{justification=centering} 
\begin{center}
{\scriptsize
\setlength{\tabcolsep}{12.1pt}
\renewcommand{\arraystretch}{1.2}
\caption{\footnotesize Properties of the voids before and after radius and density-based filtering.} 
\begin{tabular}{rrrrrrrrrrr}
\hline
Filter level & & Type & $R_{\rm lim}$  & $\rho_{\rm lim}  / \bar{\rho}$ & $N_{\rm voids}$ & $\phi_{V}$ & $\bar{R}$  & $\bar{\rho}_{\rm min} / \bar{\rho}$ & $\bar{\rho}_{\rm void} / \bar{\rho}$ \\
\hline
& & & (Mpc $h^{-1}$, min radius)  & (max density) & - & (\%) &  (Mpc $h^{-1}$) & - & -  \\
\hline
Unfiltered  & & \textit{ideal} & -  & -   &1065 &100  & 10.70 &0.13 & 1.30 & \\[-2.5pt]
& & \textit{perturbed} & - & - &2558 & 100 &  7.92 &0.36 & 1.33   \\
\hline
Soft & & \textit{ideal} & 10 & 0.2     &531 & 88  & 14.24 &0.07 & 1.04& \\[-2.5pt]
& & \textit{perturbed} & 5 & 0.33 & 1410 & 88 & 9.96 & 0.21 & 0.99  \\
\hline
Medium& & \textit{ideal} &12.5 & 0.2 &331 &      74    & 16.00 &0.06 & 0.96& \\[-2.5pt]
& & \textit{perturbed} & 5 & 0.25 & 951 & 77 &    11.01 & 0.17 & 0.90  \\
\hline
Strong& & \textit{ideal} & 15 & 0.2 & 168 &      54    &  18.32 & 0.04 & 0.86& \\[-2.5pt]
& & \textit{perturbed} & 5 & 0.2  & 614 &  62  &    11.97 & 0.14 & 0.83   \\
\hline
\label{criteria_stats}
\end{tabular}}
\caption*{\footnotesize Columns in the table: filtering level; galaxy sample type; void filtering minimum radius, $R_{\rm lim}$, and maximum density, $\rho_{\rm lim}$; number of voids, $N_{\rm voids}$; void volume fraction, $\phi_{V}$; the average void radius, $\bar{R}$; minimum density, $\bar{\rho}_{\rm min}$; and void density, $\bar{\rho}_{\rm void}$. }

\end{center}
\end{table*}

To generate the medium-filtered void sample, we first imposed a density cut on the $\rho_{\rm min}$ that eliminates all \textit{ideal} sample voids with $\rho_{\rm min} > 0.2 \bar{\rho}$. Imposing this criterion results in a $\approx 98\%$ box volume coverage in voids. Additionally, to achieve a reasonable volume occupation percentage between $\approx$ $70\%$--$90\%$, we also imposed a cut on the void radii such that we eliminated all voids in the \textit{ideal} sample with equivalent radii less than 12.5\mhp. Now voids in the \textit{ideal} sample occupy $74 \%$ of the total Illustris volume and contain 331 voids.

\begin{figure}[t]
    \includegraphics{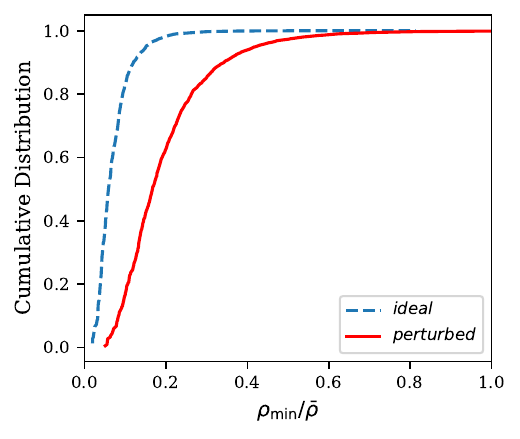}
    \caption{Cumulative volume occupation distribution depending on $\rho_{\rm min}$ for the unfiltered voids in the \textit{ideal} and \textit{perturbed} samples. 
    }
    \label{volcdf}
\end{figure}

In a similar way, we impose a density cut on the \textit{perturbed} sample voids such that $\rho_{\rm min} > 0.25 \bar{\rho}$ (the reason for a slightly higher density threshold in the \textit{perturbed} sample is because the remaining voids would only occupy $\approx$ $60 \%$ of the total volume in the box if we were to eliminate voids with $\rho_{\rm min} > 0.2 \bar{\rho}$). 
This can be better understood by looking at Fig.~\ref{volcdf}, where we show the volume filling fraction as a function of $\rho_{\rm min}$. 
We can see how in the \textit{perturbed} case, $60\%$ of the volume is occupied by voids that have $\rho_{\rm min} < 0.2  \bar{\rho}$. We also remove all voids in the \textit{perturbed} sample with a void radius less than 5\mhp. We end up with a \textit{perturbed} sample void population which occupies  $77 \%$ of the simulation box volume and contains 951 voids.

We show a similar density parameter plot for the medium-filtered void samples in the inset of Fig.~\ref{impdensparam_filt}. For these void minimum density values, the correlation between the two parameters can no longer be observed. The average $\rho_{\rm min}$ values are now $0.06 \bar{\rho}$ for voids in the \textit{ideal} and $0.17 \bar{\rho}$ for the ones in \textit{perturbed} samples, respectively.

\subsubsection{Void sizes}
\begin{figure}[t]
    \includegraphics{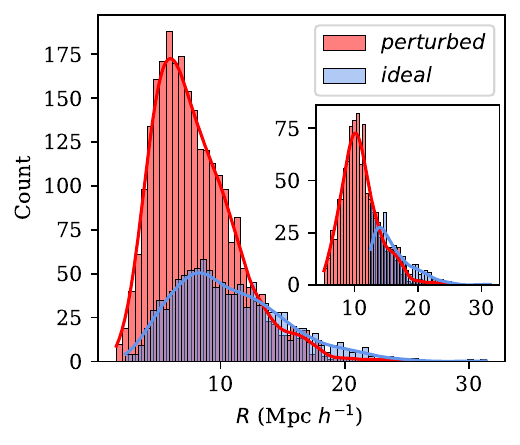}
    \caption{\textit{Main plot}: Void radius distributions for the unfiltered \textit{ideal} and \textit{perturbed} J-PAS mock samples. \textit{Inset}: same plot but for the medium-filtered voids.}
    \label{radiiimp}
\end{figure}

Next, we turned to the analysis of void sizes. As was mentioned in Section~\ref{sec:dat_and_meth}, in the first approximation, the void size is defined by Eq.~\ref{eqrad}. In Fig.~\ref{radiiimp}, we show the comparison of the void radius distributions between the \textit{ideal} and \textit{perturbed} samples considering both the unfiltered and medium-filtered voids. The unfiltered voids in the \textit{ideal} dataset span a radii interval between 2.3\mh and 31.4\mh with an average radius of 10.7\mhp. On the other hand, the unfiltered voids in the \textit{perturbed} dataset have radii between 1.6\mh and 25\mh with a mean radius of 8\mhp. The slightly smaller sizes of the voids in the \textit{perturbed} sample agree well with Fig.~\ref{voidctrimp}, where we showed how voids in the \textit{ideal} sample get fragmented into voids in the \textit{perturbed} sample, a direct consequence of the shift in the galaxy positions. If we impose the aforementioned radius and density cuts, the voids in the medium-filtered \textit{ideal} sample span radii between 12.5\mh and 31.4\mh with an average of 16\mh. In comparison, the medium-filtered \textit{perturbed} voids have radii between 5.2\mh and 25\mh with a mean radius of 11\mhp.

\subsubsection{Void density profiles}

\begin{figure*}[t]
    \centering
    \begin{minipage}[t]{0.52\textwidth}
        \centering
        \includegraphics[width=\textwidth]{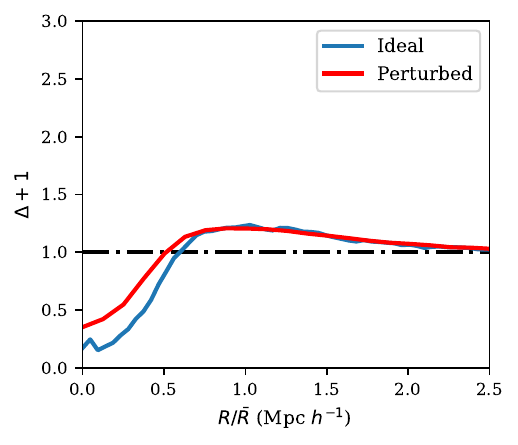}
    \end{minipage}
    \hfill
    \begin{minipage}[t]{0.465\textwidth}
        \centering
        \includegraphics[width=\textwidth]{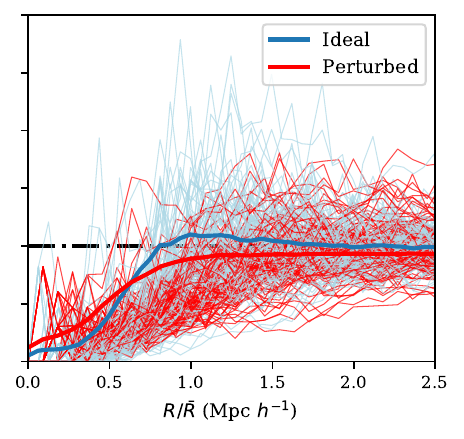}
    \end{minipage}
    \caption{Stacked radial number density profiles for the unfiltered (left) and filtered (right) \textit{ideal} and \textit{perturbed} sample voids. The horizontal line indicates the average number density of galaxies in the simulation box. The radii have been normalised to the average equivalent radii. For the filtered case, we also show the individual profiles of the 100 largest voids (thin lines, \textit{ideal} sample -- blue; \textit{perturbed} sample -- red).      }
    \label{combined_dens_prof}
\end{figure*}

Another way to probe the voids for their matter content is by computing the stacked radial number density profile $\Delta(r)$ defined as:

\begin{align}\hspace{0.3cm}
    \Delta (r) + 1 = \frac{\rho(r)}{\bar{\rho}},
\end{align}
where $\rho(r)$ is the number density of galaxies in a spherical shell of thickness ($r$, $r+$d$r$) centred on the void macrocentre and $\bar{\rho}$ is the average number density of galaxies in the simulation box.

We show the stacked radial profiles for the unfiltered \textit{ideal} and \textit{perturbed} sample voids in the left panel of Fig.~\ref{combined_dens_prof}. The radii have been normalised in each sample with the respective average void radius. Noise can be seen in the profiles -- especially in the \textit{ideal} case -- due to the low number of voids and the simplicity of the spherical model. In both samples, one can see the familiar reverse spherical top-hat profile of a void. This is characterised by a mostly flat, underdense interior that slowly rises above the average galaxy density, where the void ridge is located, and then slowly decreases towards the mean density at larger distances \citep[e.g.][]{2004MNRAS.350..517S}. The over-dense boundaries in the density profiles arise due to the ZOBOV watershed method of identifying voids. In this approach, Voronoi cells are merged to form a void until the density of adjacent cells is lower than that of the merging cells. In other words, the merging process stops when a void 'overflows' into another, creating a high-density ridge that forms the void's boundary and separates it from neighbouring voids.

By contrast, the \textit{perturbed} sample voids present slightly higher densities in their centres than the \textit{ideal} sample voids. This agrees well with the fact that \textit{perturbed} sample voids have higher $\bar{\rho}_{\rm min}$ and provide further confirmation that the redshift errors cause the ZOBOV algorithm to identify smaller and denser voids. On the other hand, the void ridges -- or the peaks of the two unfiltered profiles -- occur at around the same average density of $\approx$ 1.3 $\rho$, and both slowly decline towards the average density in the box.

In the right panel of Fig. \ref{combined_dens_prof}, we show the stacked density profiles for the medium-filtered sample together with the 100 individual profiles for the largest voids in both \textit{ideal} and \textit{perturbed} samples. The radii have been normalised with respect to the average void radii of the filtered samples. As expected, the stacked profiles peak at lower density values than their unfiltered versions. Similar to before, voids in the \textit{perturbed} sample are slightly denser in their centres than their \textit{ideal} sample counterparts. However, one can notice how the stacked void profile in the \textit{perturbed} galaxy mock does not surpass the average density in the sample. Instead, it slowly increases and plateaus at the $\bar{\rho}$ = 1. On the other hand, the voids in the \textit{ideal} sample do retain a peak above the average density, albeit a less dense one.

A plausible explanation for the filtered perturbed profile is that the redshift errors blur out overdense structures such as superclusters or filaments, thus reducing the density at the void boundary. This can be understood by looking at individual density profiles. One can notice that filtered voids in the \textit{ideal} sample have higher peaks in density in comparison with the voids in the \textit{perturbed} sample. Then why do we not see this effect in the unfiltered \textit{perturbed} sample? Despite the redshift errors blurring out the overdense structures at the void boundary, in doing so, they also raise (on average) the minimum density of a void (we can see this in Fig.~\ref{impdensparam_filt}). For a fixed $\rho_{\rm void}$, the \textit{perturbed} sample voids tend to have higher $\rho_{\rm min}$ values in comparison with the \textit{ideal} sample voids. Thus, by filtering \textit{perturbed} voids by $\rho_{\rm min}$, we also tend to eliminate voids with higher values of $\rho_{\rm void}$, which in turn will produce the stacked profile that we observe. This effect (or correlation) is less extreme in the \textit{ideal} case.

An accompanying explanation may be offered by the structures in which voids themselves are embedded. The environment around voids can provide clues about their formation and evolution. Some voids are located within large overdensities (void-in-cloud) and will end up collapsing onto themselves over time. Others are embedded in larger underdensities (void-in-void) and tend to expand and merge into larger voids. In the void-in-void case, voids are surrounded by structures having densities less than or similar to the average background density. These voids tend to expand and remain underdense. In the void-in-cloud case, the voids are surrounded by overdense structures that will cause the voids to collapse and be "squeezed out of existence" \citep{2004MNRAS.350..517S}. These two different evolutionary pathways of voids can be understood via the density profiles. We suspect that the filtering criteria we imposed on the \textit{perturbed} sample might eliminate the majority of voids embedded in overdense regions and keep only the ones that are in the void-in-void scenario. This would make sense since we are keeping only the voids that are truly underdense and these voids themselves might be surrounded by even larger underdensities. However, this explanation requires further research.

\subsection{Void recovery in J-PAS mocks}

We have seen that the photometric redshift errors affect the galaxy distribution and, on average, give rise to a larger number of smaller and denser voids in the \textit{perturbed} sample, thus affecting void identification and properties. 
The next step in our analysis was to recover as many individual \textit{ideal} sample voids as possible from the \textit{perturbed} sample, assuming that the latter is representative of what J-PAS would eventually observe.

In order to achieve this, we mapped the void volumes to a regular grid with 1~Mpc spacing using discrete Voronoi tessellation. We then compared the void volume overlap between the voids in the \textit{ideal} and \textit{perturbed} samples. In order to quantify the overlap, we considered two percentages: with respect to the \textit{ideal} sample voids, $P_{\rm i}$, and with respect to the \textit{perturbed} sample voids, $P_{\rm p}$:
\begin{align}\hspace{0.3cm}
    P_{\rm i} = \frac{V_{\rm com}}{V_{\rm i}} \label{ovi} \times 100 \%, \\
    P_{\rm p} = \frac{V_{\rm com}}{V_{\rm p}} \times 100 \%,
    \label{ovp}
\end{align}
where $V_{\mathrm{com}}$ represents the common, overlapping volume between a void in the \textit{ideal} sample and one in the \textit{perturbed} sample, $V_{\mathrm{i}}$ is the volume of a void in the \textit{ideal} sample, and $V_{\mathrm{p}}$ is the volume of a void in the \textit{perturbed} sample. For a void in the \textit{perturbed} sample to perfectly match a void in the \textit{ideal} sample (or, in other words, to fully recover an \textit{ideal} void), both $P_{\rm i}$ and $P_{\rm p}$ should be equal to $100 \%$. For example, if $P_{\rm i}$ is equal to $100 \%$, but the $P_{\rm p}$ has a different value, this means that the \textit{ideal} void is wholly engulfed in the \textit{perturbed} void, and thus does not represent the same void but just a part of it.

\subsubsection{Individual void recovery}
\label{void_ov}

\begin{figure*}[t]
    \hspace{-1.5cm}
   \centering
    \includegraphics[width=19cm]{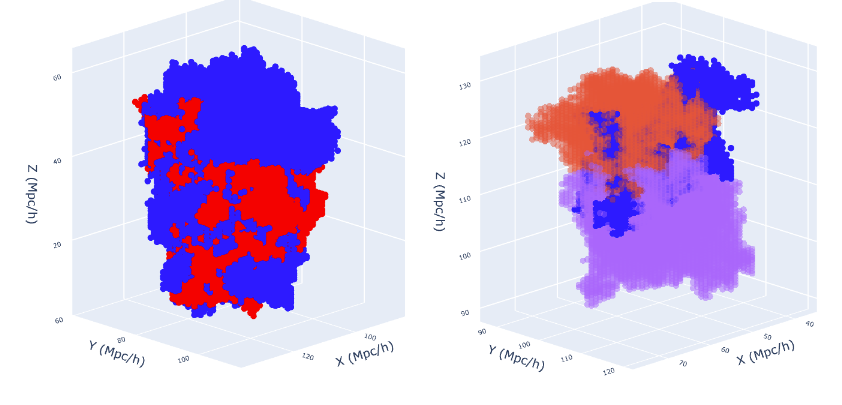}
   \caption{Example of void volume overlaps between voids in the \textit{ideal} and \textit{perturbed} samples. Left: an \textit{ideal} sample void (blue) overlapping with more than $50\%$ with a \textit{perturbed} sample void (red). Right: a different \textit{ideal} sample void (blue, solid colour) overlapping with two \textit{perturbed} sample voids, with different percentages (transparent colours).}
   \label{3dvol}
\end{figure*}

To illustrate the individual void recovery, we show two scenarios of overlap in Fig. \ref{3dvol}. 
The first scenario (left panel) shows an \textit{ideal} sample void (blue) that has a $61 \%$ percentage overlap with a \textit{perturbed} sample void (red), meaning that despite the redshift errors moving galaxies around, we can recover more than half the volume of a specific void. While this is our desired scenario, the most common case (right panel) is that an \textit{ideal} sample void (blue) is split into multiple voids in the \textit{perturbed} sample due to the redshift errors. 

We imposed a condition on both overlapping fractions to be higher than $50\%$ and will refer to these voids as "recovered voids" throughout the remaining text. 
Following this requirement, we managed to recover 53 voids, out of which 4 have more than 60$\%$ overlap, however, none surpass $70\%$.
The volumes of these voids occupy around $6.6\%$ of the box's volume. In the filtered sub-samples, the number of recovered voids decreases to 43 (soft-filtered), 29 (medium-filtered), and 17 (strong-filtered).

\begin{figure}[t]
    \centering
        \includegraphics{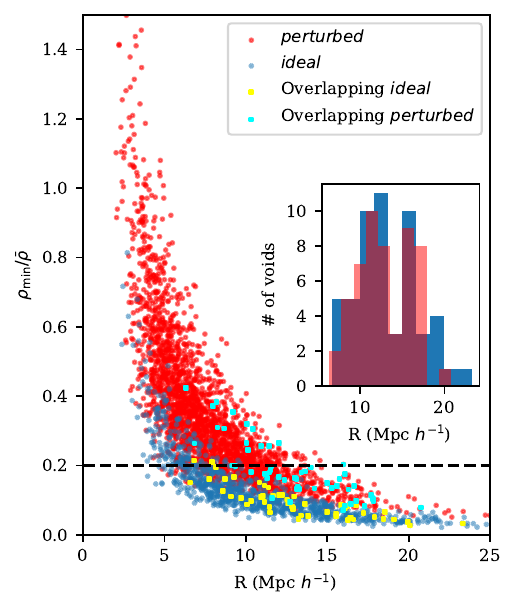}
    \caption{Dependence of the minimum void density $\rho_{\rm min}$ on the void radius $R$ for the unfiltered \textit{ideal} and \textit{perturbed} sample voids. Larger voids tend to present lower matter content. We also show the recovered voids (53 in total) that present an overlap of more than $50 \%$ with yellow (for \textit{ideal}) and cyan (for \textit{perturbed}) symbols. The inset plot shows the radii for the recovered voids. The horizontal line defines the $\rho_{\rm min} = 0.2 \bar{\rho} $ for reference.}
    \label{imp50ovnofilt}
\end{figure}

In Fig.~\ref{imp50ovnofilt}, we show the correlation between $\rho_{\rm min}$ and void radius for unfiltered \textit{ideal} and \textit{perturbed} sample voids. 
Regardless of the sample, voids with larger radii are, in general, emptier of matter, which is in agreement with the void evolution theory \citep[see e.g][]{2004MNRAS.350..517S}. 
As we have already seen, \textit{perturbed} sample voids are, on average, slightly denser and smaller than those from the \textit{ideal} sample. 
As such, the two distributions in Fig.~\ref{imp50ovnofilt} appear displaced from one another. We also show the recovered voids in \textit{ideal} sample (yellow symbols) and their \textit{perturbed} sample counterparts (cyan symbols). Their radii can be seen more clearly in the in-set histogram. In general, the two radii distributions tend to follow each other rather closely, as one would expect if the identified voids have similar volumes. 
Recovered \textit{ideal} sample voids have radii spanning an interval between 6.5\mh and 23.3\mh with an average of 13.2\mh while the overlapping \textit{perturbed} sample voids have radii in-between 6.3\mh and 20.7\mh with an average of 12.9\mhp. Looking at the average radius of the unfiltered voids in Table~\ref{criteria_stats} suggests that the identification of larger voids is potentially less affected by redshift errors, and thus, larger voids can be recovered with a higher degree of confidence than smaller ones.

Another way we tried to increase the number of recovered voids is by merging voids based on a density threshold in the unfiltered \textit{perturbed} sample (merging procedure was described in more detail in Section~\ref{sec:vide}). 
The idea behind this method is that since the \textit{perturbed} sample contains, on average, smaller voids, we could merge these voids into larger ones and thus potentially increase the void recovery.

Overlapping void recovery with merging method works in the following way:  

\begin{itemize}
  \item We choose a value for the merging threshold $\rho_{\rm threshold}$.
  \item Obtain the new \textit{perturbed} and \textit{merged} void catalogue.
  \item Compute the volume overlap percentages of the \textit{ideal} voids and the \textit{perturbed merged} voids and obtain the number of voids that have more than $50 \%$ overlap.
  \item Filter the newly obtained catalogue via density and radius criteria cuts
\end{itemize}

We tested this method for our unfiltered and medium-filtered \textit{perturbed} samples by applying 4 different density merging thresholds (normalised to the average density): 0.1, 0.2, 0.3, and 0.4. Table~\ref{overlap_stats} shows some statistics of the recovered voids for the two \textit{perturbed} samples. In general, merging voids seems to have a positive effect, increasing the number of recovered voids, albeit by a small amount. However, not every threshold can increase this number. As can be seen, thresholds of 0.1 or lower will not merge enough voids to produce a visible difference in the recovery. On the other hand, past a certain threshold ($> 0.2$) the number of recovered voids slightly increases. The act of merging will produce large, supervoids that contain within themselves smaller voids. As we increase the merging threshold, the average radii (and the box volume occupied) of the voids increase, and we end up with larger parent voids that can percolate, i.e. span the whole simulation volume. For this reason, coupled with the fact that merging voids and obtaining their volumes is computationally expensive, we do not go to thresholds higher than 0.4.

\begin{table*}[t]
\captionsetup{justification=centering} 
\begin{center}
{\scriptsize
\setlength{\tabcolsep}{5pt}
\renewcommand{\arraystretch}{1.2}
\caption{\footnotesize Statistics of unfiltered and medium-filtered \textit{perturbed} and  \textit{ideal} voids that overlap more than $50 \%$ for different merging thresholds.}
\begin{tabular}{rrrrrrrrrrr}
\hline
Filter type& & $\rho_{\rm threshold}$ & $N_{\rm voids}$ (> $50 \%$) & V$_{\rm p, box}$ (\%) & $R_{\rm p, max}$ (Mpc$\,h^{-1}$) & $\bar{R_{\rm p}}$ (Mpc$\,h^{-1}$) & V$_{\rm i, box}$ (\%) & $R_{\rm i, max}$ (Mpc$\,h^{-1}$) & $\bar{R_{\rm i}}$ (Mpc$\,h^{-1}$) \\
\hline
Unfiltered ($R_{\rm p,i~ min}$  = 6.3 / 6.5 Mpc$\,h^{-1}$)  & &  0   & 53    &  6.6 &  20.7 & 12.9 & 7.4 & 23.3 & 13.2 \\
 & & 0.1&   54   &  8.0 &    27.6 & 13.3 & 8.7 & 25.4 & 13.6\\
& &  0.2 & 69 & 14.3 &  26.8 & 14.9 & 15 & 24.2 & 15.1 \\
& &  0.3    & 69 & 14.4 &  26.8 & 14.8 & 14.2 & 24.2 & 14.9 \\
& &  0.4 &   75  & 14.9 &   26.8 & 14.9 & 15.4 & 24.2 & 14.9\\
\hline
Medium ($R_{\rm p,i~ min}$  = 11.5 / 12.5 Mpc$\,h^{-1}$)  & & 0   & 29 & 5.3   &   20.7 & 15.3 & 7.4  & 23.3 & 15.9 \\
& & 0.1&    30 & 6.7  &   27.6 & 15.9&7.5&25.4&16.5\\
& & 0.2 & 46 &   13.1 &  26.8 & 17.3& 13.8 & 24.2 & 17.7 \\
& & 0.3    & 46& 13.2 &   26.8 & 17.2 & 13.0 & 24.2 & 17.1\\
& & 0.4 &   50 &  14.4 &  26.8 & 17.2 & 13.9 & 24.2 & 17.3\\
\hline
\label{overlap_stats}
\end{tabular}}
\caption*{\footnotesize Columns show the filter type, together with the minimum void radius, $R_{\rm min}$; merging threshold, $\rho_{\rm threshold}$; the number of voids that overlap in volume with more than $50\%$, $N_{\rm voids}$ (> $50\%$); the occupied volume in the box, $V_{\rm box}$ (\%); maximum and mean void radii, $R_{\rm max}$, $\bar{R}$. The subscripts for the parameters in these columns indicate whether it's a recovered perturbed void \textit{($\mathrm{p}$)} or an ideal void \textit{($\mathrm{i}$)}.}
\end{center}
\end{table*}

\subsubsection{Void environment recovery}
Another question we tackled regards the overall void environment comparison. 
We studied the total volume of all voids in both the \textit{ideal} and \textit{perturbed} samples. 
For example, for the unfiltered \textit{ideal} and \textit{perturbed} samples, the overlap of the total void environments is $100 \%$ since the watershed constructs voids in a way that exactly divides up the whole simulation box. The picture changes, however, once we impose the filtering criteria that would only maintain the more robust voids.

Similarly to the case with individual voids, we defined two environment overlapping fractions, $V_{\mathrm{com}}/V_{\mathrm{i}}$ and $V_{\mathrm{com}}/V_{\mathrm{p}}$, except that now $V_{com}$ is the common overlapping volume of all the voids, $V_i$ and $V_p$ are the total volumes covered by the voids in the \textit{ideal} and \textit{perturbed} samples, respectively.
For medium-level filtering, we obtain values of approximately $80\%$ for both overlaps.
With respect to the simulation box volume, our common void environment occupies $61 \%$. 
Comparing this with the individual void overlap, one may interpret it in the following way: the \emph{individual voids} that VIDE identifies in the \textit{ideal} and \textit{perturbed} samples are different due to the redshift errors scrambling the spatial galaxy distribution and causing \textit{ideal} sample voids to fragment into voids of various sizes in the \textit{perturbed} sample. 
However, the overall \emph{void environment} is similar.

To assess the significance of this result, we compared it to the probability that the overlap of the void volumes of the ideal and perturbed samples is occurring completely by chance.

The simplest way is to shuffle the void grid cells for the \textit{perturbed} galaxy sample, i.e. randomise their positions. If the randomisation is uniform and independent, the resulting overlap of the void grid cells between the two samples depends only on the corresponding total volume fractions of voids. In this case, the overlap is the product of void volume fractions of both volumes $V_i \cdot V_p$.

Using the void volume fractions from the Table~\ref{criteria_stats} we obtain the following figures. For the medium-filtered voids, the common overlap volume fraction for the void environment in the data is $61\%$ of the total volume of the box and $54\%$ for the random sample. Similarly, for the soft and strong filters, the volume overlap fractions are $80\%$ and $40\%$, while for the corresponding random samples, we get $77\%$ and $31\%$, respectively. This implies that a harsher criterion will decrease the probability that the overlap occurs by chance.

\section{Discussion and conclusions}
\label{sec:disc_and_conc}

Photometric redshift errors impact the observed spatial distribution of galaxies by inserting uncertainties along the line of sight. Consequently, this affects the scientific results and our understanding of the large-scale structure of the Universe.

In this paper, we modelled the expected redshift errors of the ongoing J-PAS photometric survey in order to understand their effect on the identification and properties of voids. To tackle this issue, we created mock galaxy catalogues of J-PAS using the IllustrisTNG N-body simulation and applied realistic redshift errors estimated from the smaller precursor miniJPAS survey. We aimed to recover voids detected from an \textit{ideal} galaxy sample (without redshift errors) using a \textit{perturbed} sample (with added redshift errors). We identified voids within these mocks using VIDE, a pipeline built on the ZOBOV watershed void identification algorithm and filtered voids using density-radius criteria to reduce the number of spurious ones.

Our main results can be summarised as follows:
\begin{enumerate}
  \item Photometric redshift errors do have a strong impact on void identification and properties. Voids in the \textit{ideal} sample fragment into various-sized voids in the \textit{perturbed} sample which present, on average, denser interiors (0.36 $\bar{\rho}$ vs 0.13 $\bar{\rho}$) and smaller sizes (8.00\mh vs 10.70\mhp). These results can be seen in Figs.~\ref{radiiimp} and \ref{combined_dens_prof}, respectively. Filtering out voids based on density and radius removes the spurious, smaller and denser voids in our data. Comparing the density profiles between the filtered \textit{ideal} and \textit{perturbed} sample voids reveals that, despite their denser interiors, voids in the \textit{perturbed} sample have lower density boundaries (and lower average densities overall) in comparison with the voids in the \textit{ideal} one (right panel of Fig. \ref{combined_dens_prof}). This is a consequence of the errors blurring out surrounding overdense structures, such as clusters and filaments, which define the void boundaries.\\
  \item Recovering individual voids is difficult. We assessed the individual void recovery in the \textit{ideal} and \textit{perturbed} samples by investigating the void volume overlap between these two samples. In the unfiltered samples, we identify 53 voids (out of the total of 1065 \textit{ideal} sample voids) that overlap in volume by more than $50\%$ (which make up $6.6\%$ of the total volume of the simulation box). If we consider a medium-level filtering, this number is reduced to 29 recovered voids ($5.3\%$ volume fraction). Merging \textit{perturbed} voids by increasing the density threshold of VIDE slightly increased this number to 75 ($14.9\%$ void volume fraction) in the unfiltered sample and 50 ($14.4\%$) in the medium-filtered sample. In general, merging voids has a positive effect on void recovery, albeit a small one. Furthermore, we find that larger voids are less affected by the redshift errors and, thus, more likely to be recovered then smaller voids (Fig. \ref{imp50ovnofilt}).\\
  \item General void environment recovery is somewhat better than the recovery of individual voids. We also looked at the overall void environment overlap between the medium-filtered samples. This is defined as the total volume occupied by voids. We found that the environment overlaps with a percentage of $80 \%$ and occupies $61\%$ volume of the simulation box for the medium-filtered voids. However, note that given the large filling fraction of voids, even a random labelling would recover a significant fraction of the void grid cells, in our case $54\%$ of the simulation box volume. Increasing the strength of the filter does increase the difference in percentages such that our strongest filtering will result in a $40\%$ box volume occupation of the void environment overlap in the real samples and $31\%$ in the random ones.
\end{enumerate}

\cite{2024ApJ...962...58C} has investigated properties of voids and void galaxies in the IllustrisTNG300-1 simulation using a spherical void-finding algorithm. They found 5078 voids at a snapshot of $z = 0$. Their smallest void has a radius of 2.5\mhp, comparable with the smallest void in our unfiltered \textit{ideal} sample, while their largest void has a radius of 24.7\mhp. The median void radius found by the authors is 4.4\mhp. Overall, we found larger voids in both \textit{ideal} and \textit{perturbed} samples, which might be a consequence of the magnitude cuts that we impose, essentially decreasing the number density of galaxies and thus increasing the average void radius \citep[see e.g ][for how tracer population impacts void properties]{2015MNRAS.454..889N, 2017MNRAS.470.4434P, 2024MNRAS.529.4325B}. Furthermore, it can also be a consequence of the Voronoi tessellation used by the ZOBOV algorithm to identify voids. Since ZOBOV tessellates the whole galaxy distribution in the box, voids will end up having overdense boundaries made out of filaments and walls. This, in turn, will make the volumes of the voids -- and thus the equivalent radii -- larger than the ones identified by a spherical void finder. The radius is also differently defined: VIDE defines the void radius as the radius of a sphere that has the same volume as the volume of a void (which in turn is defined as the sum of the volumes of the Voronoi cells). In contrast, the spherical void finder constructs spheres around void centres with radii such that the average void density is less than $\bar{\rho}$. Their luminous density profile is also comparable to our unfiltered \textit{ideal} sample, with an underdense interior of $\approx 0.2 \bar{\rho}$. However, their profile does peak at a smaller radius since their voids are smaller on average. Another study by \cite{2023MNRAS.518.3095D} also found void sizes between 7--11\mh when investigating galactic spins with respect to void centres in TNG300-1 simulation.

Regarding the individual void recovery, \cite{2017MNRAS.465..746S} identified voids from a spectroscopic sample and from an equivalent photometric one by projecting galaxies into 2D slices and identifying voids within the 2D density field. The authors conclude that voids with sizes much larger than the photo-$z$ error should not be affected by galaxy scattering. The median value of our photo-$z$ error is $\approx$ 2\mhp, almost 6 times smaller than the average radii of the recovered voids. Interestingly, the authors mention that the structure on scales comparable to the photo-$z$ errors will end up being erased. We tend to find that the smearing of galaxies will actually cause our \textit{ideal} sample voids to end up being split into different \textit{perturbed} sample voids, which, on average, present smaller radii. That being said, the stacked density profile of the filtered \textit{perturbed} sample voids does peak at a lower density than the \textit{ideal} sample one (right panel of Fig. \ref{combined_dens_prof}), suggesting that the overdense structures, such as filaments and clusters bounding these voids, are smeared out.

There are some definite limitations of our work. First of all, the number of miniJPAS galaxies used for estimating the photometric redshift errors is very small (79), especially in comparison with the number of galaxies in IllustrisTNG (358186). As more J-PAS data will be collected, this limitation can be overcome in future work. For example, a larger sample of observed galaxies might enable the selection of a subsample of sources with lower errors. Furthermore, J-PAS will provide additional information for void detection via weak-lensing analysis \citep{2016MNRAS.455.3367G, 2023MNRAS.525...91P}.  Secondly, the watershed ZOBOV algorithm proved not to be ideal for recovering voids based on volume overlap since it is extremely sensitive to galaxy tracers.

The recovered voids could potentially be used in studying the environmental effect that low-density regions have on the galaxy properties, while the overall void population statistics could be used in cosmological applications. Therefore, our future work will focus on improving the recovery of voids through different methods. For example, one might select only luminous red galaxies since they have lower photo-$z$ errors or use clustering algorithms to trace galaxy groups instead of individual galaxies, thus averaging out the redshift uncertainties. Alternatively, a different void finder can be considered, such as one based on the Delaunay tessellation, since this would capture the density field more accurately via interpolation.

\begin{acknowledgements}
We would like to thank the anonymous referee for the 
thorough report and useful comments which helped us to improve the manuscript. We acknowledge the support by the Estonian Ministry of Education and Research (grant TK202), Estonian Research Council grant (PRG2172, PRG1006, PSG700, PRG2159) and the European Union's Horizon Europe research and innovation programme (EXCOSM, grant No. 101159513). This work is based on observations made with the JST/T250 telescope and JPCam at the Observatorio Astrofísico de Javalambre (OAJ), in Teruel, owned, managed, and operated by the Centro de Estudios de Física del Cosmos de Aragón (CEFCA). We acknowledge the OAJ Data Processing and Archiving Department (DPAD) for reducing and calibrating the OAJ data used in this work. Funding for the J-PAS Project has been provided by the Governments of Spain and Aragón through the Fondo de Inversión de Teruel, European FEDER funding and the Spanish Ministry of Science, Innovation and Universities, and by the Brazilian agencies FINEP, FAPESP, FAPERJ and by the National Observatory of Brazil with additional funding also provided by the University of Tartu and by the J-PAS Chinese Astronomical Consortium. VM thanks CNPq (Brazil) and FAPES (Brazil) for partial financial support. This paper has gone through an internal review by the J-PAS collaboration.
\end{acknowledgements}

%


\bibliographystyle{aa} 
\bibliography{ref} 

\begin{thebibliography}{89}
\expandafter\ifx\csname natexlab\endcsname\relax\def\natexlab#1{#1}\fi

\bibitem[{{Astropy Collaboration} {et~al.}(2022){Astropy Collaboration},
  {Price-Whelan}, {Lim}, {Earl}, {Starkman}, {Bradley}, {Shupe}, {Patil},
  {Corrales}, {Brasseur}, {N{"o}the}, {Donath}, {Tollerud}, {Morris},
  {Ginsburg}, {Vaher}, {Weaver}, {Tocknell}, {Jamieson}, {van Kerkwijk},
  {Robitaille}, {Merry}, {Bachetti}, {G{"u}nther}, {Aldcroft},
  {Alvarado-Montes}, {Archibald}, {B{'o}di}, {Bapat}, {Barentsen}, {Baz{'a}n},
  {Biswas}, {Boquien}, {Burke}, {Cara}, {Cara}, {Conroy}, {Conseil}, {Craig},
  {Cross}, {Cruz}, {D'Eugenio}, {Dencheva}, {Devillepoix}, {Dietrich},
  {Eigenbrot}, {Erben}, {Ferreira}, {Foreman-Mackey}, {Fox}, {Freij}, {Garg},
  {Geda}, {Glattly}, {Gondhalekar}, {Gordon}, {Grant}, {Greenfield}, {Groener},
  {Guest}, {Gurovich}, {Handberg}, {Hart}, {Hatfield-Dodds}, {Homeier},
  {Hosseinzadeh}, {Jenness}, {Jones}, {Joseph}, {Kalmbach}, {Karamehmetoglu},
  {Ka{l}uszy{'n}ski}, {Kelley}, {Kern}, {Kerzendorf}, {Koch}, {Kulumani},
  {Lee}, {Ly}, {Ma}, {MacBride}, {Maljaars}, {Muna}, {Murphy}, {Norman},
  {O'Steen}, {Oman}, {Pacifici}, {Pascual}, {Pascual-Granado}, {Patil},
  {Perren}, {Pickering}, {Rastogi}, {Roulston}, {Ryan}, {Rykoff}, {Sabater},
  {Sakurikar}, {Salgado}, {Sanghi}, {Saunders}, {Savchenko}, {Schwardt},
  {Seifert-Eckert}, {Shih}, {Jain}, {Shukla}, {Sick}, {Simpson},
  {Singanamalla}, {Singer}, {Singhal}, {Sinha}, {Sip{H{o}}cz}, {Spitler},
  {Stansby}, {Streicher}, {{{S}}umak}, {Swinbank}, {Taranu}, {Tewary},
  {Tremblay}, {Val-Borro}, {Van Kooten}, {Vasovi{'c}}, {Verma}, {de Miranda
  Cardoso}, {Williams}, {Wilson}, {Winkel}, {Wood-Vasey}, {Xue}, {Yoachim},
  {Zhang}, {Zonca}, \& {Astropy Project Contributors}}]{astropy:2022}
{Astropy Collaboration}, {Price-Whelan}, A.~M., {Lim}, P.~L., {et~al.} 2022,
  apj, 935, 167

\bibitem[{{Astropy Collaboration} {et~al.}(2018){Astropy Collaboration},
  {Price-Whelan}, {Sip{\H{o}}cz}, {G{\"u}nther}, {Lim}, {Crawford}, {Conseil},
  {Shupe}, {Craig}, {Dencheva}, {Ginsburg}, {Vand erPlas}, {Bradley},
  {P{\'e}rez-Su{\'a}rez}, {de Val-Borro}, {Aldcroft}, {Cruz}, {Robitaille},
  {Tollerud}, {Ardelean}, {Babej}, {Bach}, {Bachetti}, {Bakanov}, {Bamford},
  {Barentsen}, {Barmby}, {Baumbach}, {Berry}, {Biscani}, {Boquien}, {Bostroem},
  {Bouma}, {Brammer}, {Bray}, {Breytenbach}, {Buddelmeijer}, {Burke},
  {Calderone}, {Cano Rodr{\'\i}guez}, {Cara}, {Cardoso}, {Cheedella}, {Copin},
  {Corrales}, {Crichton}, {D'Avella}, {Deil}, {Depagne}, {Dietrich}, {Donath},
  {Droettboom}, {Earl}, {Erben}, {Fabbro}, {Ferreira}, {Finethy}, {Fox},
  {Garrison}, {Gibbons}, {Goldstein}, {Gommers}, {Greco}, {Greenfield},
  {Groener}, {Grollier}, {Hagen}, {Hirst}, {Homeier}, {Horton}, {Hosseinzadeh},
  {Hu}, {Hunkeler}, {Ivezi{\'c}}, {Jain}, {Jenness}, {Kanarek}, {Kendrew},
  {Kern}, {Kerzendorf}, {Khvalko}, {King}, {Kirkby}, {Kulkarni}, {Kumar},
  {Lee}, {Lenz}, {Littlefair}, {Ma}, {Macleod}, {Mastropietro}, {McCully},
  {Montagnac}, {Morris}, {Mueller}, {Mumford}, {Muna}, {Murphy}, {Nelson},
  {Nguyen}, {Ninan}, {N{\"o}the}, {Ogaz}, {Oh}, {Parejko}, {Parley}, {Pascual},
  {Patil}, {Patil}, {Plunkett}, {Prochaska}, {Rastogi}, {Reddy Janga},
  {Sabater}, {Sakurikar}, {Seifert}, {Sherbert}, {Sherwood-Taylor}, {Shih},
  {Sick}, {Silbiger}, {Singanamalla}, {Singer}, {Sladen}, {Sooley},
  {Sornarajah}, {Streicher}, {Teuben}, {Thomas}, {Tremblay}, {Turner},
  {Terr{\'o}n}, {van Kerkwijk}, {de la Vega}, {Watkins}, {Weaver}, {Whitmore},
  {Woillez}, {Zabalza}, \& {Astropy Contributors}}]{astropy:2018}
{Astropy Collaboration}, {Price-Whelan}, A.~M., {Sip{\H{o}}cz}, B.~M., {et~al.}
  2018, \aj, 156, 123

\bibitem[{{Astropy Collaboration} {et~al.}(2013){Astropy Collaboration},
  {Robitaille}, {Tollerud}, {Greenfield}, {Droettboom}, {Bray}, {Aldcroft},
  {Davis}, {Ginsburg}, {Price-Whelan}, {Kerzendorf}, {Conley}, {Crighton},
  {Barbary}, {Muna}, {Ferguson}, {Grollier}, {Parikh}, {Nair}, {Unther},
  {Deil}, {Woillez}, {Conseil}, {Kramer}, {Turner}, {Singer}, {Fox}, {Weaver},
  {Zabalza}, {Edwards}, {Azalee Bostroem}, {Burke}, {Casey}, {Crawford},
  {Dencheva}, {Ely}, {Jenness}, {Labrie}, {Lim}, {Pierfederici}, {Pontzen},
  {Ptak}, {Refsdal}, {Servillat}, \& {Streicher}}]{astropy:2013}
{Astropy Collaboration}, {Robitaille}, T.~P., {Tollerud}, E.~J., {et~al.} 2013,
  \aap, 558, A33

\bibitem[{Benitez {et~al.}(2014)Benitez, Dupke, Moles, Sodre, Cenarro,
  Marin-Franch, Taylor, Cristobal, Fernández-Soto, Oliveira, Cepa, Abramo,
  Alcaniz, Overzier, Hernández-Monteagudo, Alfaro, Kanaan, Carvano, Reis, \&
  Valdivielso}]{jpas_article}
Benitez, N., Dupke, R., Moles, M., {et~al.} 2014

\bibitem[{{Bermejo} {et~al.}(2024){Bermejo}, {Wilding}, {van de Weygaert},
  {Jones}, {Vegter}, \& {Efstathiou}}]{2024MNRAS.529.4325B}
{Bermejo}, R., {Wilding}, G., {van de Weygaert}, R., {et~al.} 2024, \mnras,
  529, 4325

\bibitem[{{Bertschinger}(1985)}]{1985ApJS...58....1B}
{Bertschinger}, E. 1985, \apjs, 58, 1

\bibitem[{{Bertschinger}(1998)}]{1998ARA&A..36..599B}
{Bertschinger}, E. 1998, \araa, 36, 599

\bibitem[{Beygu {et~al.}(2016)Beygu, Peletier, van der Hulst, Jarrett,
  Kreckel, Weygaert, van Gorkom, \& Aragon-Calvo}]{10.1093/mnras/stw2362}
Beygu, B., Peletier, R.~F., van der Hulst, J.~M., {et~al.} 2016, Monthly
  Notices of the Royal Astronomical Society, 464, 666

\bibitem[{{Blumenthal} {et~al.}(1992){Blumenthal}, {da Costa}, {Goldwirth},
  {Lecar}, \& {Piran}}]{1992ApJ...388..234B}
{Blumenthal}, G.~R., {da Costa}, L.~N., {Goldwirth}, D.~S., {Lecar}, M., \&
  {Piran}, T. 1992, \apj, 388, 234

\bibitem[{{Bond} {et~al.}(1991){Bond}, {Cole}, {Efstathiou}, \&
  {Kaiser}}]{1991ApJ...379..440B}
{Bond}, J.~R., {Cole}, S., {Efstathiou}, G., \& {Kaiser}, N. 1991, \apj, 379,
  440

\bibitem[{{Bonoli} {et~al.}(2021){Bonoli}, {Mar{\'\i}n-Franch}, {Varela},
  {V{\'a}zquez Rami{\'o}}, {Abramo}, {Cenarro}, {Dupke}, {V{\'\i}lchez},
  {Crist{\'o}bal-Hornillos}, {Gonz{\'a}lez Delgado},
  {Hern{\'a}ndez-Monteagudo}, {L{\'o}pez-Sanjuan}, {Muniesa}, {Civera},
  {Ederoclite}, {Hern{\'a}n-Caballero}, {Marra}, {Baqui}, {Cortesi},
  {Cypriano}, {Daflon}, {de Amorim}, {D{\'\i}az-Garc{\'\i}a}, {Diego},
  {Mart{\'\i}nez-Solaeche}, {P{\'e}rez}, {Placco}, {Prada}, {Queiroz},
  {Alcaniz}, {Alvarez-Candal}, {Cepa}, {Maroto}, {Roig}, {Siffert}, {Taylor},
  {Benitez}, {Moles}, {Sodr{\'e}}, {Carneiro}, {Mendes de Oliveira}, {Abdalla},
  {Angulo}, {Aparicio Resco}, {Balaguera-Antol{\'\i}nez}, {Ballesteros},
  {Brito-Silva}, {Broadhurst}, {Carrasco}, {Castro}, {Cid Fernandes}, {Coelho},
  {de Melo}, {Doubrawa}, {Fernandez-Soto}, {Ferrari}, {Finoguenov},
  {Garc{\'\i}a-Benito}, {Iglesias-P{\'a}ramo}, {Jim{\'e}nez-Teja}, {Kitaura},
  {Laur}, {Lopes}, {Lucatelli}, {Mart{\'\i}nez}, {Maturi}, {Overzier},
  {Pigozzo}, {Quartin}, {Rodr{\'\i}guez-Mart{\'\i}n}, {Salzano}, {Tamm},
  {Tempel}, {Umetsu}, {Valdivielso}, {von Marttens}, {Zitrin},
  {D{\'\i}az-Mart{\'\i}n}, {L{\'o}pez-Alegre}, {L{\'o}pez-Sainz},
  {Yanes-D{\'\i}az}, {Rueda-Teruel}, {Rueda-Teruel}, {Abril Iba{\~n}ez}, {L
  Ant{\'o}n Bravo}, {Bello Ferrer}, {Bielsa}, {Casino}, {Castillo}, {Chueca},
  {Cuesta}, {Garzar{\'a}n Calderaro}, {Iglesias-Marzoa}, {{\'I}niguez},
  {Lamadrid Gutierrez}, {Lopez-Martinez}, {Lozano-P{\'e}rez}, {Ma{\'\i}cas
  Sacrist{\'a}n}, {Molina-Ib{\'a}{\~n}ez}, {Moreno-Signes}, {Rodr{\'\i}guez
  Llano}, {Royo Navarro}, {Tilve Rua}, {Andrade}, {Alfaro}, {Akras},
  {Arnalte-Mur}, {Ascaso}, {Barbosa}, {Beltr{\'a}n Jim{\'e}nez}, {Benetti},
  {Bengaly}, {Bernui}, {Blanco-Pillado}, {Borges Fernandes}, {Bregman},
  {Bruzual}, {Calderone}, {Carvano}, {Casarini}, {Chaves-Montero},
  {Chies-Santos}, {Coutinho de Carvalho}, {Dimauro}, {Duarte Puertas},
  {Figueruelo}, {Gonz{\'a}lez-Serrano}, {Guerrero}, {Gurung-L{\'o}pez},
  {Herranz}, {Huertas-Company}, {Irwin}, {Izquierdo-Villalba}, {Kanaan},
  {Kehrig}, {Kirkpatrick}, {Lim}, {Lopes}, {Lopes de Oliveira},
  {Marcos-Caballero}, {Mart{\'\i}nez-Delgado}, {Mart{\'\i}nez-Gonz{\'a}lez},
  {Mart{\'\i}nez-Somonte}, {Oliveira}, {Orsi}, {Penna-Lima}, {Reis}, {Spinoso},
  {Tsujikawa}, {Vielva}, {Vitorelli}, {Xia}, {Yuan}, {Arroyo-Polonio},
  {Dantas}, {Galarza}, {Gon{\c{c}}alves}, {Gon{\c{c}}alves}, {Gonzalez},
  {Gonzalez}, {Greisel}, {Jim{\'e}nez-Esteban}, {Landim}, {Lazzaro}, {Magris},
  {Monteiro-Oliveira}, {Pereira}, {Rebou{\c{c}}as}, {Rodriguez-Espinosa},
  {Santos da Costa}, \& {Telles}}]{2021A&A...653A..31B}
{Bonoli}, S., {Mar{\'\i}n-Franch}, A., {Varela}, J., {et~al.} 2021, \aap, 653,
  A31

\bibitem[{{Boquien} {et~al.}(2019){Boquien}, {Burgarella}, {Roehlly}, {Buat},
  {Ciesla}, {Corre}, {Inoue}, \& {Salas}}]{2019A&A...622A.103B}
{Boquien}, M., {Burgarella}, D., {Roehlly}, Y., {et~al.} 2019, \aap, 622, A103

\bibitem[{Bromley \& Geller(2024)}]{bromley2024cosmologyvoids}
Bromley, B.~C. \& Geller, M.~J. 2024, Cosmology with voids

\bibitem[{{Buchert} {et~al.}(2000){Buchert}, {Kerscher}, \&
  {Sicka}}]{2000PhRvD..62d3525B}
{Buchert}, T., {Kerscher}, M., \& {Sicka}, C. 2000, \prd, 62, 043525

\bibitem[{{Carfora} \& {Familiari}(2024)}]{2024arXiv240104293C}
{Carfora}, M. \& {Familiari}, F. 2024, arXiv e-prints, arXiv:2401.04293

\bibitem[{{Clampitt} \& {Jain}(2015)}]{2015MNRAS.454.3357C}
{Clampitt}, J. \& {Jain}, B. 2015, \mnras, 454, 3357

\bibitem[{{Colberg} {et~al.}(2008){Colberg}, {Pearce}, {Foster}, {Platen},
  {Brunino}, {Neyrinck}, {Basilakos}, {Fairall}, {Feldman}, {Gottl{\"o}ber},
  {Hahn}, {Hoyle}, {M{\"u}ller}, {Nelson}, {Plionis}, {Porciani}, {Shandarin},
  {Vogeley}, \& {van de Weygaert}}]{2008MNRAS.387..933C}
{Colberg}, J.~M., {Pearce}, F., {Foster}, C., {et~al.} 2008, \mnras, 387, 933

\bibitem[{{Colberg} {et~al.}(2005){Colberg}, {Sheth}, {Diaferio}, {Gao}, \&
  {Yoshida}}]{2005MNRAS.360..216C}
{Colberg}, J.~M., {Sheth}, R.~K., {Diaferio}, A., {Gao}, L., \& {Yoshida}, N.
  2005, \mnras, 360, 216

\bibitem[{Colless {et~al.}(2003)Colless, Peterson, Jackson, Peacock, Cole,
  Norberg, Baldry, Baugh, Bland-Hawthorn, Bridges, Cannon, Collins, Couch,
  Cross, Dalton, Propris, Driver, Efstathiou, Ellis, Frenk, Glazebrook, Lahav,
  Lewis, Lumsden, Maddox, Madgwick, Sutherland, \& Taylor}]{colless20032df}
Colless, M., Peterson, B.~A., Jackson, C., {et~al.} 2003, The 2dF Galaxy
  Redshift Survey: Final Data Release

\bibitem[{{Conrado} {et~al.}(2024){Conrado}, {Gonz{\'a}lez Delgado},
  {Garc{\'\i}a-Benito}, {P{\'e}rez}, {Verley}, {Ruiz-Lara},
  {S{\'a}nchez-Menguiano}, {Duarte Puertas}, {Jim{\'e}nez},
  {Dom{\'\i}nguez-G{\'o}mez}, {Espada}, {Argudo-Fern{\'a}ndez},
  {Alc{\'a}zar-Laynez}, {Bl{\'a}zquez-Calero}, {Bidaran}, {Zurita}, {Peletier},
  {Torres-R{\'\i}os}, {Florido}, {Rodr{\'\i}guez Mart{\'\i}nez}, {del
  Moral-Castro}, {van de Weygaert}, {Falc{\'o}n-Barroso}, {Lugo-Aranda},
  {S{\'a}nchez}, {van der Hulst}, {Courtois}, {Ferr{\'e}-Mateu},
  {S{\'a}nchez-Bl{\'a}zquez}, {Rom{\'a}n}, \& {Aceituno}}]{2024A&A...687A..98C}
{Conrado}, A.~M., {Gonz{\'a}lez Delgado}, R.~M., {Garc{\'\i}a-Benito}, R.,
  {et~al.} 2024, \aap, 687, A98

\bibitem[{{Contarini} {et~al.}(2024){Contarini}, {Pisani}, {Hamaus}, {Marulli},
  {Moscardini}, \& {Baldi}}]{2024A&A...682A..20C}
{Contarini}, S., {Pisani}, A., {Hamaus}, N., {et~al.} 2024, \aap, 682, A20

\bibitem[{{Contarini} {et~al.}(2022){Contarini}, {Verza}, {Pisani}, {Hamaus},
  {Sahl{\'e}n}, {Carbone}, {Dusini}, {Marulli}, {Moscardini}, {Renzi},
  {Sirignano}, {Stanco}, {Aubert}, {Bonici}, {Castignani}, {Courtois},
  {Escoffier}, {Guinet}, {Kovacs}, {Lavaux}, {Massara}, {Nadathur}, {Pollina},
  {Ronconi}, {Ruppin}, {Sakr}, {Veropalumbo}, {Wandelt}, {Amara}, {Auricchio},
  {Baldi}, {Bonino}, {Branchini}, {Brescia}, {Brinchmann}, {Camera},
  {Capobianco}, {Carretero}, {Castellano}, {Cavuoti}, {Cledassou}, {Congedo},
  {Conselice}, {Conversi}, {Copin}, {Corcione}, {Courbin}, {Cropper}, {Da
  Silva}, {Degaudenzi}, {Dubath}, {Duncan}, {Dupac}, {Ealet}, {Farrens},
  {Ferriol}, {Fosalba}, {Frailis}, {Franceschi}, {Garilli}, {Gillard},
  {Gillis}, {Giocoli}, {Grazian}, {Grupp}, {Guzzo}, {Haugan}, {Holmes},
  {Hormuth}, {Jahnke}, {K{\"u}mmel}, {Kermiche}, {Kiessling}, {Kilbinger},
  {Kunz}, {Kurki-Suonio}, {Laureijs}, {Ligori}, {Lilje}, {Lloro}, {Maiorano},
  {Mansutti}, {Marggraf}, {Markovic}, {Massey}, {Melchior}, {Meneghetti},
  {Meylan}, {Moresco}, {Munari}, {Niemi}, {Padilla}, {Paltani}, {Pasian},
  {Pedersen}, {Percival}, {Pettorino}, {Pires}, {Polenta}, {Poncet}, {Popa},
  {Pozzetti}, {Raison}, {Rhodes}, {Rossetti}, {Saglia}, {Sartoris},
  {Schneider}, {Secroun}, {Seidel}, {Sirri}, {Surace}, {Tallada-Cresp{\'\i}},
  {Taylor}, {Tereno}, {Toledo-Moreo}, {Torradeflot}, {Valentijn}, {Valenziano},
  {Wang}, {Weller}, {Zamorani}, {Zoubian}, {Andreon}, {Maino}, \&
  {Mei}}]{2022A&A...667A.162C}
{Contarini}, S., {Verza}, G., {Pisani}, A., {et~al.} 2022, \aap, 667, A162

\bibitem[{{Curtis} {et~al.}(2024){Curtis}, {McDonough}, \&
  {Brainerd}}]{2024ApJ...962...58C}
{Curtis}, O., {McDonough}, B., \& {Brainerd}, T.~G. 2024, \apj, 962, 58

\bibitem[{{D{\'a}vila-Kurb{\'a}n} {et~al.}(2023){D{\'a}vila-Kurb{\'a}n},
  {Lares}, \& {Lambas}}]{2023MNRAS.518.3095D}
{D{\'a}vila-Kurb{\'a}n}, F., {Lares}, M., \& {Lambas}, D.~G. 2023, \mnras, 518,
  3095

\bibitem[{{de Lapparent} {et~al.}(1986){de Lapparent}, {Geller}, \&
  {Huchra}}]{1986ApJ...302L...1D}
{de Lapparent}, V., {Geller}, M.~J., \& {Huchra}, J.~P. 1986, \apjl, 302, L1

\bibitem[{{Douglass} {et~al.}(2022){Douglass}, {Veyrat}, {Zaidouni}, \&
  {BenZvi}}]{2022AAS...24013923D}
{Douglass}, K., {Veyrat}, D., {Zaidouni}, F., \& {BenZvi}, S. 2022, in American
  Astronomical Society Meeting Abstracts, Vol.~54, American Astronomical
  Society Meeting Abstracts, 139.23

\bibitem[{{Dressler}(1980)}]{1980ApJ...236..351D}
{Dressler}, A. 1980, \apj, 236, 351

\bibitem[{{Einasto} {et~al.}(2020){Einasto}, {Deshev}, {Tenjes},
  {Hein{\"a}m{\"a}ki}, {Tempel}, {Juhan Liivam{\"a}gi}, {Einasto}, {Lietzen},
  {Tuvikene}, \& {Chon}}]{2020A&A...641A.172E}
{Einasto}, M., {Deshev}, B., {Tenjes}, P., {et~al.} 2020, \aap, 641, A172

\bibitem[{{Einasto} {et~al.}(2022){Einasto}, {Kipper}, {Tenjes}, {Einasto},
  {Tempel}, \& {Liivam{\"a}gi}}]{2022A&A...668A..69E}
{Einasto}, M., {Kipper}, R., {Tenjes}, P., {et~al.} 2022, \aap, 668, A69

\bibitem[{{Einasto} {et~al.}(2021){Einasto}, {Kipper}, {Tenjes}, {Lietzen},
  {Tempel}, {Liivam{\"a}gi}, {Einasto}, {Tamm}, {Hein{\"a}m{\"a}ki}, \&
  {Nurmi}}]{2021A&A...649A..51E}
{Einasto}, M., {Kipper}, R., {Tenjes}, P., {et~al.} 2021, \aap, 649, A51

\bibitem[{{Fang} {et~al.}(2019){Fang}, {Hamaus}, {Jain}, {Pandey}, {Pollina},
  {S{\'a}nchez}, {Kov{\'a}cs}, {Chang}, {Carretero}, {Castander}, {Choi},
  {Crocce}, {DeRose}, {Fosalba}, {Gatti}, {Gazta{\~n}aga}, {Gruen}, {Hartley},
  {Hoyle}, {MacCrann}, {Prat}, {Rau}, {Rykoff}, {Samuroff}, {Sheldon},
  {Troxel}, {Vielzeuf}, {Zuntz}, {Annis}, {Avila}, {Bertin}, {Brooks}, {Burke},
  {Carnero Rosell}, {Carrasco Kind}, {Cawthon}, {da Costa}, {De Vicente},
  {Desai}, {Diehl}, {Dietrich}, {Doel}, {Everett}, {Evrard}, {Flaugher},
  {Frieman}, {Garc{\'\i}a-Bellido}, {Gerdes}, {Gruendl}, {Gutierrez},
  {Hollowood}, {James}, {Jarvis}, {Kuropatkin}, {Lahav}, {Maia}, {Marshall},
  {Melchior}, {Menanteau}, {Miquel}, {Palmese}, {Plazas}, {Romer}, {Roodman},
  {Sanchez}, {Serrano}, {Sevilla-Noarbe}, {Smith}, {Soares-Santos}, {Sobreira},
  {Suchyta}, {Swanson}, {Tarle}, {Thomas}, {Vikram}, {Walker}, {Weller}, \&
  {DES Collaboration}}]{2019MNRAS.490.3573F}
{Fang}, Y., {Hamaus}, N., {Jain}, B., {et~al.} 2019, \mnras, 490, 3573

\bibitem[{Ganeshaiah Veena {et~al.}(2019)Ganeshaiah Veena, Cautun, Tempel,
  van de Weygaert, \& Frenk}]{Ganeshaiah_Veena_2019}
Ganeshaiah Veena, P., Cautun, M., Tempel, E., van de Weygaert, R., \& Frenk,
  C.~S. 2019, Monthly Notices of the Royal Astronomical Society, 487,
  1607–1625

\bibitem[{{Genel} {et~al.}(2014){Genel}, {Vogelsberger}, {Springel}, {Sijacki},
  {Nelson}, {Snyder}, {Rodriguez-Gomez}, {Torrey}, \&
  {Hernquist}}]{2014MNRAS.445..175G}
{Genel}, S., {Vogelsberger}, M., {Springel}, V., {et~al.} 2014, \mnras, 445,
  175

\bibitem[{{Gonz{\'a}lez Delgado} {et~al.}(2021){Gonz{\'a}lez Delgado},
  {D{\'\i}az-Garc{\'\i}a}, {de Amorim}, {Bruzual}, {Cid Fernandes},
  {P{\'e}rez}, {Bonoli}, {Cenarro}, {Coelho}, {Cortesi}, {Garc{\'\i}a-Benito},
  {L{\'o}pez Fern{\'a}ndez}, {Mart{\'\i}nez-Solaeche},
  {Rodr{\'\i}guez-Mart{\'\i}n}, {Magris}, {Mej{\'\i}a-Narvaez}, {Brito-Silva},
  {Abramo}, {Diego}, {Dupke}, {Hern{\'a}n-Caballero},
  {Hern{\'a}ndez-Monteagudo}, {L{\'o}pez-Sanjuan}, {Mar{\'\i}n-Franch},
  {Marra}, {Moles}, {Montero-Dorta}, {Queiroz}, {Sodr{\'e}}, {Varela},
  {V{\'a}zquez Rami{\'o}}, {V{\'\i}lchez}, {Baqui}, {Ben{\'\i}tez},
  {Crist{\'o}bal-Hornillos}, {Ederoclite}, {Mendes de Oliveira}, {Civera},
  {Muniesa}, {Taylor}, {Tempel}, \& {J-PAS
  Collaboration}}]{2021A&A...649A..79G}
{Gonz{\'a}lez Delgado}, R.~M., {D{\'\i}az-Garc{\'\i}a}, L.~A., {de Amorim}, A.,
  {et~al.} 2021, \aap, 649, A79

\bibitem[{Grogin \& Geller(1999)}]{Grogin_1999}
Grogin, N.~A. \& Geller, M.~J. 1999, The Astronomical Journal, 118, 2561–2580

\bibitem[{{Gruen} {et~al.}(2016){Gruen}, {Friedrich}, {Amara}, {Bacon},
  {Bonnett}, {Hartley}, {Jain}, {Jarvis}, {Kacprzak}, {Krause}, {Mana}, {Rozo},
  {Rykoff}, {Seitz}, {Sheldon}, {Troxel}, {Vikram}, {Abbott}, {Abdalla},
  {Allam}, {Armstrong}, {Banerji}, {Bauer}, {Becker}, {Benoit-L{\'e}vy},
  {Bernstein}, {Bernstein}, {Bertin}, {Bridle}, {Brooks}, {Buckley-Geer},
  {Burke}, {Capozzi}, {Carnero Rosell}, {Carrasco Kind}, {Carretero}, {Crocce},
  {Cunha}, {D'Andrea}, {da Costa}, {DePoy}, {Desai}, {Diehl}, {Dietrich},
  {Doel}, {Eifler}, {Neto}, {Fernandez}, {Flaugher}, {Fosalba}, {Frieman},
  {Gerdes}, {Gruendl}, {Gutierrez}, {Honscheid}, {James}, {Kuehn},
  {Kuropatkin}, {Lahav}, {Li}, {Lima}, {Maia}, {March}, {Martini}, {Melchior},
  {Miller}, {Miquel}, {Mohr}, {Nord}, {Ogando}, {Plazas}, {Reil}, {Romer},
  {Roodman}, {Sako}, {Sanchez}, {Scarpine}, {Schubnell}, {Sevilla-Noarbe},
  {Smith}, {Soares-Santos}, {Sobreira}, {Suchyta}, {Swanson}, {Tarle},
  {Thaler}, {Thomas}, {Walker}, {Wechsler}, {Weller}, {Zhang}, \&
  {Zuntz}}]{2016MNRAS.455.3367G}
{Gruen}, D., {Friedrich}, O., {Amara}, A., {et~al.} 2016, \mnras, 455, 3367

\bibitem[{Hamaus {et~al.}(2016)Hamaus, Pisani, Sutter, Lavaux, Escoffier,
  Wandelt, \& Weller}]{Hamaus_2016}
Hamaus, N., Pisani, A., Sutter, P., {et~al.} 2016, Physical Review Letters, 117

\bibitem[{{Hamaus} {et~al.}(2015){Hamaus}, {Sutter}, {Lavaux}, \&
  {Wandelt}}]{2015JCAP...11..036H}
{Hamaus}, N., {Sutter}, P.~M., {Lavaux}, G., \& {Wandelt}, B.~D. 2015, \jcap,
  2015, 036

\bibitem[{{Hawken} {et~al.}(2020){Hawken}, {Aubert}, {Pisani}, {Cousinou},
  {Escoffier}, {Nadathur}, {Rossi}, \& {Schneider}}]{2020JCAP...06..012H}
{Hawken}, A.~J., {Aubert}, M., {Pisani}, A., {et~al.} 2020, \jcap, 2020, 012

\bibitem[{Hellwing {et~al.}(2021)Hellwing, Cautun, van~de Weygaert, \&
  Jones}]{Hellwing_2021}
Hellwing, W.~A., Cautun, M., van~de Weygaert, R., \& Jones, B.~T. 2021,
  Physical Review D, 103

\bibitem[{{Hern{\'a}n-Caballero} {et~al.}(2024){Hern{\'a}n-Caballero},
  {Akhlaghi}, {L{\'o}pez-Sanjuan}, {V{\'a}zquez Rami{\'o}}, {Laur}, {Varela},
  {Civera}, {Muniesa}, {Finoguenov}, {Fern{\'a}ndez-Ontiveros}, {Dom{\'\i}nguez
  S{\'a}nchez}, {Chaves-Montero}, {Fern{\'a}ndez-Soto}, {Lumbreras-Calle},
  {D{\'\i}az-Garc{\'\i}a}, {del Pino}, {Gonz{\'a}lez Delgado},
  {Hern{\'a}ndez-Monteagudo}, {Coelho}, {Jim{\'e}nez-Teja}, {Lopes}, {Marra},
  {Tempel}, {V{\'\i}lchez}, {Abramo}, {Alcaniz}, {Ben{\'\i}tez}, {Bonoli},
  {Carneiro}, {Cenarro}, {Crist{\'o}bal-Hornillos}, {Dupke}, {Ederoclite},
  {Mar{\'\i}n-Franch}, {Mendes de Oliveira}, {Moles}, {Sodr{\'e}}, \&
  {Taylor}}]{2024A&A...684A..61H}
{Hern{\'a}n-Caballero}, A., {Akhlaghi}, M., {L{\'o}pez-Sanjuan}, C., {et~al.}
  2024, \aap, 684, A61

\bibitem[{{Hern{\'a}n-Caballero} {et~al.}(2021){Hern{\'a}n-Caballero},
  {Varela}, {L{\'o}pez-Sanjuan}, {Muniesa}, {Civera}, {Chaves-Montero},
  {D{\'\i}az-Garc{\'\i}a}, {Laur}, {Hern{\'a}ndez-Monteagudo}, {Abramo},
  {Angulo}, {Crist{\'o}bal-Hornillos}, {Gonz{\'a}lez Delgado}, {Greisel},
  {Orsi}, {Queiroz}, {Sobral}, {Tamm}, {Tempel}, {V{\'a}zquez-Rami{\'o}},
  {Alcaniz}, {Ben{\'\i}tez}, {Bonoli}, {Carneiro}, {Cenarro}, {Dupke},
  {Ederoclite}, {Mar{\'\i}n-Franch}, {Mendes de Oliveira}, {Moles},
  {Sodr{\'e}}, {Taylor}, {Cypriano}, \&
  {Mart{\'\i}nez-Solaeche}}]{2021A&A...654A.101H}
{Hern{\'a}n-Caballero}, A., {Varela}, J., {L{\'o}pez-Sanjuan}, C., {et~al.}
  2021, \aap, 654, A101

\bibitem[{{Hoffman} \& {Shaham}(1982)}]{1982ApJ...262L..23H}
{Hoffman}, Y. \& {Shaham}, J. 1982, \apjl, 262, L23

\bibitem[{{Hoyle} \& {Vogeley}(2002)}]{2002ApJ...566..641H}
{Hoyle}, F. \& {Vogeley}, M.~S. 2002, \apj, 566, 641

\bibitem[{Hoyle \& Vogeley(2004)}]{Hoyle_2004}
Hoyle, F. \& Vogeley, M.~S. 2004, The Astrophysical Journal, 607, 751–764

\bibitem[{{Hoyle} {et~al.}(2012){Hoyle}, {Vogeley}, \&
  {Pan}}]{2012MNRAS.426.3041H}
{Hoyle}, F., {Vogeley}, M.~S., \& {Pan}, D. 2012, \mnras, 426, 3041

\bibitem[{{Icke}(1984)}]{1984MNRAS.206P...1I}
{Icke}, V. 1984, \mnras, 206, 1P

\bibitem[{{J{\~o}eveer} {et~al.}(1978){J{\~o}eveer}, {Einasto}, \&
  {Tago}}]{1978MNRAS.185..357J}
{J{\~o}eveer}, M., {Einasto}, J., \& {Tago}, E. 1978, \mnras, 185, 357

\bibitem[{{Jaber} {et~al.}(2024){Jaber}, {Peper}, {Hellwing},
  {Arag{\'o}n-Calvo}, \& {Valenzuela}}]{2024MNRAS.527.4087J}
{Jaber}, M., {Peper}, M., {Hellwing}, W.~A., {Arag{\'o}n-Calvo}, M.~A., \&
  {Valenzuela}, O. 2024, \mnras, 527, 4087

\bibitem[{Jennings {et~al.}(2013)Jennings, Li, \& Hu}]{Jennings_2013}
Jennings, E., Li, Y., \& Hu, W. 2013, Monthly Notices of the Royal Astronomical
  Society, 434, 2167–2181

\bibitem[{{Kirshner} {et~al.}(1987){Kirshner}, {Oemler}, {Schechter}, \&
  {Shectman}}]{1987ApJ...314..493K}
{Kirshner}, R.~P., {Oemler}, Augustus, J., {Schechter}, P.~L., \& {Shectman},
  S.~A. 1987, \apj, 314, 493

\bibitem[{Kitaura {et~al.}(2016)Kitaura, Chuang, Liang, Zhao, Tao,
  Rodríguez-Torres, Eisenstein, Gil-Marín, Kneib, McBride, Percival, Ross,
  Sánchez, Tinker, Tojeiro, Vargas-Magana, \& Zhao}]{Kitaura_2016}
Kitaura, F.-S., Chuang, C.-H., Liang, Y., {et~al.} 2016, Physical Review
  Letters, 116

\bibitem[{Kreckel {et~al.}(2012)Kreckel, Platen, Aragón-Calvo, van Gorkom,
  van~de Weygaert, van~der Hulst, \& Beygu}]{Kreckel_2012}
Kreckel, K., Platen, E., Aragón-Calvo, M.~A., {et~al.} 2012, The Astronomical
  Journal, 144, 16

\bibitem[{Kreisch {et~al.}(2019)Kreisch, Pisani, Carbone, Liu, Hawken, Massara,
  Spergel, \& Wandelt}]{10.1093/mnras/stz1944}
Kreisch, C.~D., Pisani, A., Carbone, C., {et~al.} 2019, Monthly Notices of the
  Royal Astronomical Society, 488, 4413

\bibitem[{{Kreisch} {et~al.}(2022){Kreisch}, {Pisani}, {Villaescusa-Navarro},
  {Spergel}, {Wandelt}, {Hamaus}, \& {Bayer}}]{2022ApJ...935..100K}
{Kreisch}, C.~D., {Pisani}, A., {Villaescusa-Navarro}, F., {et~al.} 2022, \apj,
  935, 100

\bibitem[{{Lacey} \& {Cole}(1993)}]{1993MNRAS.262..627L}
{Lacey}, C. \& {Cole}, S. 1993, \mnras, 262, 627

\bibitem[{{Laur} {et~al.}(2022){Laur}, {Tempel}, {Tamm}, {Kipper},
  {Liivam{\"a}gi}, {Hern{\'a}n-Caballero}, {Muru}, {Chaves-Montero},
  {D{\'\i}az-Garc{\'\i}a}, {Turner}, {Tuvikene}, {Queiroz}, {Bom},
  {Fern{\'a}ndez-Ontiveros}, {Gonz{\'a}lez Delgado}, {Civera}, {Abramo},
  {Alcaniz}, {Ben{\'\i}tez}, {Bonoli}, {Carneiro}, {Cenarro},
  {Crist{\'o}bal-Hornillos}, {Dupke}, {Ederoclite}, {L{\'o}pez-Sanjuan},
  {Mar{\'\i}n-Franch}, {de Oliveira}, {Moles}, {Sodr{\'e}}, {Taylor}, {Varela},
  \& {V{\'a}zquez Rami{\'o}}}]{2022A&A...668A...8L}
{Laur}, J., {Tempel}, E., {Tamm}, A., {et~al.} 2022, \aap, 668, A8

\bibitem[{{Lavaux} \& {Wandelt}(2010)}]{2010MNRAS.403.1392L}
{Lavaux}, G. \& {Wandelt}, B.~D. 2010, \mnras, 403, 1392

\bibitem[{Liddle \& Lyth(1993)}]{LIDDLE19931}
Liddle, A.~R. \& Lyth, D.~H. 1993, Physics Reports, 231, 1

\bibitem[{{Marinacci} {et~al.}(2018){Marinacci}, {Vogelsberger}, {Pakmor},
  {Torrey}, {Springel}, {Hernquist}, {Nelson}, {Weinberger}, {Pillepich},
  {Naiman}, \& {Genel}}]{2018MNRAS.480.5113M}
{Marinacci}, F., {Vogelsberger}, M., {Pakmor}, R., {et~al.} 2018, \mnras, 480,
  5113

\bibitem[{Maturi {et~al.}(2023)Maturi, Finoguenov, Lopes, González~Delgado,
  Dupke, Cypriano, Carrasco, Diego, Penna-Lima, Doubrawa, Vílchez, Moscardini,
  Marra, Bonoli, Rodríguez-Martín, Zitrin, Márquez, Hernán-Caballero,
  Jiménez-Teja, Abramo, Alcaniz, Benitez, Carneiro, Cenarro,
  Cristóbal-Hornillos, Ederoclite, López-Sanjuan, Marín-Franch, Mendes~de
  Oliveira, Moles, Sodré~Jr, Taylor, Varela, Vázquez~Ramió, \&
  Fernández-Ontiveros}]{Maturi_2023}
Maturi, M., Finoguenov, A., Lopes, P. A.~A., {et~al.} 2023, \aap, 678, A145

\bibitem[{{Nadathur} \& {Hotchkiss}(2014)}]{2014MNRAS.440.1248N}
{Nadathur}, S. \& {Hotchkiss}, S. 2014, \mnras, 440, 1248

\bibitem[{{Nadathur} \& {Hotchkiss}(2015)}]{2015MNRAS.454..889N}
{Nadathur}, S. \& {Hotchkiss}, S. 2015, \mnras, 454, 889

\bibitem[{{Naiman} {et~al.}(2018){Naiman}, {Pillepich}, {Springel},
  {Ramirez-Ruiz}, {Torrey}, {Vogelsberger}, {Pakmor}, {Nelson}, {Marinacci},
  {Hernquist}, {Weinberger}, \& {Genel}}]{2018MNRAS.477.1206N}
{Naiman}, J.~P., {Pillepich}, A., {Springel}, V., {et~al.} 2018, \mnras, 477,
  1206

\bibitem[{{Nelson} {et~al.}(2018){Nelson}, {Pillepich}, {Springel},
  {Weinberger}, {Hernquist}, {Pakmor}, {Genel}, {Torrey}, {Vogelsberger},
  {Kauffmann}, {Marinacci}, \& {Naiman}}]{2018MNRAS.475..624N}
{Nelson}, D., {Pillepich}, A., {Springel}, V., {et~al.} 2018, \mnras, 475, 624

\bibitem[{{Neyrinck}(2008)}]{2008MNRAS.386.2101N}
{Neyrinck}, M.~C. 2008, \mnras, 386, 2101

\bibitem[{{Padilla} {et~al.}(2005){Padilla}, {Ceccarelli}, \&
  {Lambas}}]{2005MNRAS.363..977P}
{Padilla}, N.~D., {Ceccarelli}, L., \& {Lambas}, D.~G. 2005, \mnras, 363, 977

\bibitem[{{Paillas} {et~al.}(2017){Paillas}, {Lagos}, {Padilla}, {Tissera},
  {Helly}, \& {Schaller}}]{2017MNRAS.470.4434P}
{Paillas}, E., {Lagos}, C. D.~P., {Padilla}, N., {et~al.} 2017, \mnras, 470,
  4434

\bibitem[{Pan {et~al.}(2012)Pan, Vogeley, Hoyle, Choi, \& Park}]{Pan_2012}
Pan, D.~C., Vogeley, M.~S., Hoyle, F., Choi, Y.-Y., \& Park, C. 2012, Monthly
  Notices of the Royal Astronomical Society, 421, 926–934

\bibitem[{{Peper} \& {Roukema}(2021)}]{2021MNRAS.505.1223P}
{Peper}, M. \& {Roukema}, B.~F. 2021, \mnras, 505, 1223

\bibitem[{{Peper} {et~al.}(2023){Peper}, {Roukema}, \&
  {Bolejko}}]{2023MNRAS.525...91P}
{Peper}, M., {Roukema}, B.~F., \& {Bolejko}, K. 2023, \mnras, 525, 91

\bibitem[{{Pillepich} {et~al.}(2018){Pillepich}, {Nelson}, {Hernquist},
  {Springel}, {Pakmor}, {Torrey}, {Weinberger}, {Genel}, {Naiman}, {Marinacci},
  \& {Vogelsberger}}]{2018MNRAS.475..648P}
{Pillepich}, A., {Nelson}, D., {Hernquist}, L., {et~al.} 2018, \mnras, 475, 648

\bibitem[{{Pisani} {et~al.}(2019){Pisani}, {Massara}, {Spergel}, {Alonso},
  {Baker}, {Cai}, {Cautun}, {Davies}, {Demchenko}, {Dor{\'e}}, {Goulding},
  {Habouzit}, {Hamaus}, {Hawken}, {Hirata}, {Ho}, {Jain}, {Kreisch}, {Marulli},
  {Padilla}, {Pollina}, {Sahl{\'e}n}, {Sheth}, {Somerville}, {Szapudi}, {van de
  Weygaert}, {Villaescusa-Navarro}, {Wandelt}, \& {Wang}}]{2019BAAS...51c..40P}
{Pisani}, A., {Massara}, E., {Spergel}, D.~N., {et~al.} 2019, \baas, 51, 40

\bibitem[{{Planck Collaboration} {et~al.}(2016){Planck Collaboration}, {Ade},
  {Aghanim}, {Arnaud}, {Ashdown}, {Aumont}, {Baccigalupi}, {Banday},
  {Barreiro}, {Bartlett}, {Bartolo}, {Battaner}, {Battye}, {Benabed},
  {Beno{\^\i}t}, {Benoit-L{\'e}vy}, {Bernard}, {Bersanelli}, {Bielewicz},
  {Bock}, {Bonaldi}, {Bonavera}, {Bond}, {Borrill}, {Bouchet}, {Boulanger},
  {Bucher}, {Burigana}, {Butler}, {Calabrese}, {Cardoso}, {Catalano},
  {Challinor}, {Chamballu}, {Chary}, {Chiang}, {Chluba}, {Christensen},
  {Church}, {Clements}, {Colombi}, {Colombo}, {Combet}, {Coulais}, {Crill},
  {Curto}, {Cuttaia}, {Danese}, {Davies}, {Davis}, {de Bernardis}, {de Rosa},
  {de Zotti}, {Delabrouille}, {D{\'e}sert}, {Di Valentino}, {Dickinson},
  {Diego}, {Dolag}, {Dole}, {Donzelli}, {Dor{\'e}}, {Douspis}, {Ducout},
  {Dunkley}, {Dupac}, {Efstathiou}, {Elsner}, {En{\ss}lin}, {Eriksen},
  {Farhang}, {Fergusson}, {Finelli}, {Forni}, {Frailis}, {Fraisse},
  {Franceschi}, {Frejsel}, {Galeotta}, {Galli}, {Ganga}, {Gauthier}, {Gerbino},
  {Ghosh}, {Giard}, {Giraud-H{\'e}raud}, {Giusarma}, {Gjerl{\o}w},
  {Gonz{\'a}lez-Nuevo}, {G{\'o}rski}, {Gratton}, {Gregorio}, {Gruppuso},
  {Gudmundsson}, {Hamann}, {Hansen}, {Hanson}, {Harrison}, {Helou},
  {Henrot-Versill{\'e}}, {Hern{\'a}ndez-Monteagudo}, {Herranz}, {Hildebrandt},
  {Hivon}, {Hobson}, {Holmes}, {Hornstrup}, {Hovest}, {Huang}, {Huffenberger},
  {Hurier}, {Jaffe}, {Jaffe}, {Jones}, {Juvela}, {Keih{\"a}nen}, {Keskitalo},
  {Kisner}, {Kneissl}, {Knoche}, {Knox}, {Kunz}, {Kurki-Suonio}, {Lagache},
  {L{\"a}hteenm{\"a}ki}, {Lamarre}, {Lasenby}, {Lattanzi}, {Lawrence}, {Leahy},
  {Leonardi}, {Lesgourgues}, {Levrier}, {Lewis}, {Liguori}, {Lilje},
  {Linden-V{\o}rnle}, {L{\'o}pez-Caniego}, {Lubin}, {Mac{\'\i}as-P{\'e}rez},
  {Maggio}, {Maino}, {Mandolesi}, {Mangilli}, {Marchini}, {Maris}, {Martin},
  {Martinelli}, {Mart{\'\i}nez-Gonz{\'a}lez}, {Masi}, {Matarrese}, {McGehee},
  {Meinhold}, {Melchiorri}, {Melin}, {Mendes}, {Mennella}, {Migliaccio},
  {Millea}, {Mitra}, {Miville-Desch{\^e}nes}, {Moneti}, {Montier}, {Morgante},
  {Mortlock}, {Moss}, {Munshi}, {Murphy}, {Naselsky}, {Nati}, {Natoli},
  {Netterfield}, {N{\o}rgaard-Nielsen}, {Noviello}, {Novikov}, {Novikov},
  {Oxborrow}, {Paci}, {Pagano}, {Pajot}, {Paladini}, {Paoletti}, {Partridge},
  {Pasian}, {Patanchon}, {Pearson}, {Perdereau}, {Perotto}, {Perrotta},
  {Pettorino}, {Piacentini}, {Piat}, {Pierpaoli}, {Pietrobon}, {Plaszczynski},
  {Pointecouteau}, {Polenta}, {Popa}, {Pratt}, {Pr{\'e}zeau}, {Prunet},
  {Puget}, {Rachen}, {Reach}, {Rebolo}, {Reinecke}, {Remazeilles}, {Renault},
  {Renzi}, {Ristorcelli}, {Rocha}, {Rosset}, {Rossetti}, {Roudier},
  {Rouill{\'e} d'Orfeuil}, {Rowan-Robinson}, {Rubi{\~n}o-Mart{\'\i}n},
  {Rusholme}, {Said}, {Salvatelli}, {Salvati}, {Sandri}, {Santos},
  {Savelainen}, {Savini}, {Scott}, {Seiffert}, {Serra}, {Shellard}, {Spencer},
  {Spinelli}, {Stolyarov}, {Stompor}, {Sudiwala}, {Sunyaev}, {Sutton},
  {Suur-Uski}, {Sygnet}, {Tauber}, {Terenzi}, {Toffolatti}, {Tomasi},
  {Tristram}, {Trombetti}, {Tucci}, {Tuovinen}, {T{\"u}rler}, {Umana},
  {Valenziano}, {Valiviita}, {Van Tent}, {Vielva}, {Villa}, {Wade}, {Wandelt},
  {Wehus}, {White}, {White}, {Wilkinson}, {Yvon}, {Zacchei}, \&
  {Zonca}}]{2016A&A...594A..13P}
{Planck Collaboration}, {Ade}, P.~A.~R., {Aghanim}, N., {et~al.} 2016, \aap,
  594, A13

\bibitem[{{Platen} {et~al.}(2007){Platen}, {van de Weygaert}, \&
  {Jones}}]{2007MNRAS.380..551P}
{Platen}, E., {van de Weygaert}, R., \& {Jones}, B. J.~T. 2007, \mnras, 380,
  551

\bibitem[{{Press} \& {Schechter}(1974)}]{1974ApJ...187..425P}
{Press}, W.~H. \& {Schechter}, P. 1974, \apj, 187, 425

\bibitem[{{R{\"a}s{\"a}nen}(2004)}]{2004JCAP...02..003R}
{R{\"a}s{\"a}nen}, S. 2004, \jcap, 2004, 003

\bibitem[{Rosas-Guevara {et~al.}(2022)Rosas-Guevara, Tissera, Lagos, Paillas,
  \& Padilla}]{Rosas_Guevara_2022}
Rosas-Guevara, Y., Tissera, P., Lagos, C. d.~P., Paillas, E., \& Padilla, N.
  2022, Monthly Notices of the Royal Astronomical Society, 517, 712–731

\bibitem[{{Roukema}(2018)}]{2018A&A...610A..51R}
{Roukema}, B.~F. 2018, \aap, 610, A51

\bibitem[{{S{\'a}nchez} {et~al.}(2017){S{\'a}nchez}, {Clampitt}, {Kovacs},
  {Jain}, {Garc{\'\i}a-Bellido}, {Nadathur}, {Gruen}, {Hamaus}, {Huterer},
  {Vielzeuf}, {Amara}, {Bonnett}, {DeRose}, {Hartley}, {Jarvis}, {Lahav},
  {Miquel}, {Rozo}, {Rykoff}, {Sheldon}, {Wechsler}, {Zuntz}, {Abbott},
  {Abdalla}, {Annis}, {Benoit-L{\'e}vy}, {Bernstein}, {Bernstein}, {Bertin},
  {Brooks}, {Buckley-Geer}, {Carnero Rosell}, {Carrasco Kind}, {Carretero},
  {Crocce}, {Cunha}, {D'Andrea}, {da Costa}, {Desai}, {Diehl}, {Dietrich},
  {Doel}, {Evrard}, {Fausti Neto}, {Flaugher}, {Fosalba}, {Frieman},
  {Gaztanaga}, {Gruendl}, {Gutierrez}, {Honscheid}, {James}, {Krause}, {Kuehn},
  {Lima}, {Maia}, {Marshall}, {Melchior}, {Plazas}, {Reil}, {Romer}, {Sanchez},
  {Schubnell}, {Sevilla-Noarbe}, {Smith}, {Soares-Santos}, {Sobreira},
  {Suchyta}, {Tarle}, {Thomas}, {Walker}, {Weller}, \& {DES
  Collaboration}}]{2017MNRAS.465..746S}
{S{\'a}nchez}, C., {Clampitt}, J., {Kovacs}, A., {et~al.} 2017, \mnras, 465,
  746

\bibitem[{Schuster {et~al.}(2019)Schuster, Hamaus, Pisani, Carbone, Kreisch,
  Pollina, \& Weller}]{Schuster_2019}
Schuster, N., Hamaus, N., Pisani, A., {et~al.} 2019, Journal of Cosmology and
  Astroparticle Physics, 2019, 055

\bibitem[{{Sheth} \& {van de Weygaert}(2004)}]{2004MNRAS.350..517S}
{Sheth}, R.~K. \& {van de Weygaert}, R. 2004, \mnras, 350, 517

\bibitem[{{Springel} {et~al.}(2018){Springel}, {Pakmor}, {Pillepich},
  {Weinberger}, {Nelson}, {Hernquist}, {Vogelsberger}, {Genel}, {Torrey},
  {Marinacci}, \& {Naiman}}]{2018MNRAS.475..676S}
{Springel}, V., {Pakmor}, R., {Pillepich}, A., {et~al.} 2018, \mnras, 475, 676

\bibitem[{{Sutter} {et~al.}(2014{\natexlab{a}}){Sutter}, {Lavaux}, {Hamaus},
  {Pisani}, {Wandelt}, {Warren}, {Villaescusa-Navarro}, {Zivick}, {Mao}, \&
  {Thompson}}]{2014ascl.soft07014S}
{Sutter}, P.~M., {Lavaux}, G., {Hamaus}, N., {et~al.} 2014{\natexlab{a}},
  {VIDE: The Void IDentification and Examination toolkit}, Astrophysics Source
  Code Library, record ascl:1407.014

\bibitem[{{Sutter} {et~al.}(2014{\natexlab{b}}){Sutter}, {Pisani}, {Wandelt},
  \& {Weinberg}}]{2014MNRAS.443.2983S}
{Sutter}, P.~M., {Pisani}, A., {Wandelt}, B.~D., \& {Weinberg}, D.~H.
  2014{\natexlab{b}}, \mnras, 443, 2983

\bibitem[{{Tamone} {et~al.}(2023){Tamone}, {Zhao}, {Forero-S{\'a}nchez},
  {Variu}, {Chuang}, {Kitaura}, {Kneib}, \& {Tao}}]{2023MNRAS.526.2889T}
{Tamone}, A., {Zhao}, C., {Forero-S{\'a}nchez}, D., {et~al.} 2023, \mnras, 526,
  2889

\bibitem[{{Tempel} \& {Tamm}(2015)}]{2015A&A...576L...5T}
{Tempel}, E. \& {Tamm}, A. 2015, \aap, 576, L5

\bibitem[{{van de Weygaert} \& {van Kampen}(1993)}]{1993MNRAS.263..481V}
{van de Weygaert}, R. \& {van Kampen}, E. 1993, \mnras, 263, 481

\bibitem[{{Verweg} {et~al.}(2023){Verweg}, {Jones}, \& {van de
  Weygaert}}]{2023arXiv231019451V}
{Verweg}, D. B.~H., {Jones}, B. J.~T., \& {van de Weygaert}, R. 2023, arXiv
  e-prints, arXiv:2310.19451

\end{thebibliography}

\end{document}